\documentclass[a4paper, amsfonts, amssymb, amsmath, reprint, showkeys, nofootinbib, twoside, floatfix]{revtex4-1}
\usepackage[english]{babel}
\usepackage{braket}

\usepackage{amsthm}
\usepackage{mathtools}
\usepackage{physics}
\usepackage{xcolor}
\usepackage{graphicx}
\usepackage[left=23mm,right=13mm,top=35mm,columnsep=15pt]{geometry} 
\usepackage{adjustbox}
\usepackage{placeins}
\usepackage[T1]{fontenc}
\usepackage{lipsum}
\usepackage{csquotes}
\usepackage{amsmath,amssymb}
\usepackage{graphicx}
\usepackage{tabularx,booktabs}
\usepackage{tikz}
\usepackage{float}
\usetikzlibrary{quantikz}
\usepackage[utf8]{inputenc}
\usepackage[colorinlistoftodos, color=green!40, prependcaption]{todonotes}
\usepackage[pdftex, pdftitle={Article}, pdfauthor={Author}]{hyperref}
\usepackage[ruled,lined]{algorithm2e}
\usepackage{algpseudocode}
\usepackage{hyperref}

\usepackage{caption}

\usetikzlibrary{shapes.geometric, arrows}
\usetikzlibrary{automata,positioning}

\tikzstyle{startstop} = [rectangle, rounded corners, minimum width=3cm, minimum height=1cm, text centered, draw=black, fill=blue!20]
\tikzstyle{io} = [trapezium, trapezium left angle=70, trapezium right angle=110, minimum width=3cm, minimum height=1cm, text centered, draw=black, fill=blue!30]
\tikzstyle{process} = [rectangle, minimum width=3cm, minimum height=1cm, text centered, draw=black, fill=blue!20]
\tikzstyle{decision} = [diamond, minimum width=3cm, minimum height=1cm, text centered, draw=black, fill= green!30]
\tikzstyle{arrow} = [thick, ->, >=stealth]

\begin{document}

\title{An evolving objective function for improved \\ variational quantum optimisation}
\author{Ioannis Kolotouros}
    \email{i.kolotouros@sms.ed.ac.uk}
    \affiliation{University of Edinburgh, School of Informatics, EH8 9AB Edinburgh, United Kingdom}
\author{Petros Wallden}
    \email{petros.wallden@ed.ac.uk}
    \affiliation{University of Edinburgh, School of Informatics, EH8 9AB Edinburgh, United Kingdom}
\date{\today}

\begin{abstract}
A promising approach to useful computational quantum advantage is to use variational quantum algorithms for optimisation problems. Crucial for the performance of these algorithms is to ensure that the algorithm converges with high probability to a near-optimal solution in a small time. In Barkoutsos et al [Quantum 2020] an alternative class of objective functions, called Conditional Value-at-Risk (CVaR), was introduced and it was shown that they perform better than standard objective functions. Here we extend that work by introducing an evolving objective function, which we call Ascending-CVaR and that can be used for any optimisation problem. We test our proposed objective function, in an emulation environment, using as case-studies three different optimisation problems: Max-Cut, Number Partitioning and Portfolio Optimisation. We examine multiple instances of different sizes and analyse the performance using the Variational Quantum Eigensolver (VQE) with hardware-efficient ansatz and the Quantum Approximate Optimization Algorithm (QAOA). We show that Ascending-CVaR in all cases performs better than standard objective functions or the ``constant'' CVaR of Barkoutsos et al [Quantum 2020] and that it can be used as a heuristic for avoiding sub-optimal minima. Our proposal achieves higher overlap with the ideal state in all problems, whether we consider easy or hard instances -- on average it gives up to ten times greater overlap at Portfolio Optimisation and Number Partitioning, while it gives an 80\% improvement at Max-Cut. In the hard instances we consider, for the number partitioning problem, standard objective functions fail to find the correct solution in almost all cases, CVaR finds the correct solution at 60\% of the cases, while Ascending-CVaR finds the correct solution in 95\% of the cases. 
\end{abstract}

\maketitle

\section{Introduction}

We have recently entered the era where quantum computers have scaled up, from small proof-of-principle devices to devices that are beyond the classical simulation limit opening the prospect for providing computational speed-ups. However, we are still very far from the point that large fault tolerant quantum computers are developed. Our period has been termed as Noisy Intermediate Scale Quantum (NISQ) device era \cite{preskill2018quantum} and refers to the time that the existing devices vary from $\approx 50$ qubits of Google's quantum-advantage\footnote{Also known as ``quantum computational supremacy''.} experiment \cite{arute2019quantum} to devices with $O(1000)$ qubits that are anticipated in a horizon of five to ten years.

There are two paths forward for quantum computing. The ``long-term'' path requires to intensify the efforts (theoretical and experimental) to overcome existing barriers and truly scale up these devices to the large fault-tolerant regime. The ``near-term'' one, is to determine if and how these  NISQ devices can be used directly and offer advantage for problems of practical importance. A promising approach in the latter path, is the use of hybrid quantum-classical algorithms. A leading class of candidate algorithms, both due the possible importance of the applications and the promise it shows, is the class of variational quantum algorithms for optimisation problems.

One can divide variational quantum algorithms (see more details \hyperref[preliminaries]{II}]) into three main steps. The first step is to map the targeted problem to the mathematical task that these algorithms are designed to solve, which is the search for the ground state energy of a Hamiltonian\footnote{Mathematically this is simply evaluating the smallest eigenvalue of a Hermitian matrix.}. The second step, is a method to estimate the energy of a quantum state, given a (polynomial in the size of the input) number of copies. Finally, the third step consists of a parameterised family of quantum states (``ansatz'') and a classical optimiser that given the above tools, outputs efficiently an approximation of the ground state energy. This is done by finding the choice of parameters that lead to the quantum state that has the smallest energy.

The success of the algorithms depend on all those steps and extensive research on improving each of them exists, indicatively, \cite{egger2020warm} used warm-starting to improve QAOA on low depth, \cite{bravyi2019obstacles} improved QAOA by introducing a non-local version which outperformed classical QAOA on 3-regular graphs, \cite{skolik2021layerwise, nakanishi2020sequential} introduced different procedures on how to optimise the variational parameters and \cite{wauters2020reinforcement} used reinforcement learning to assist the classical optimisation. 

What we focus in this contribution is the third part, and specifically on how to use the measurement outcomes performed in estimating the energy of a quantum state to (i) accelerate the speed and (ii) improve the accuracy that the classical optimiser finds an (approximation of the) ground state and thus solves the problem optimally. Prior to our work, \cite{li2020quantum} inspired by statistical physics, used a Gibbs objective function to improve the performance. Minimising the \emph{infidelity} between the parameterised state and a target state \cite{benedetti2019generative, cheng2018information} appears to be another promising approach.

For classical optimisation problems, the solution (ground state) is one of the computational basis quantum states. Preparing a quantum state that has big overlap with that state is sufficient to give a good and quick approximation of the ground state. For example, if one can achieve a constant but possibly small overlap with the correct solution, it is guaranteed after sampling this state a constant number of times to obtain at least one sample of the true ground state. In \cite{barkoutsos2020improving}, the authors used this idea, and instead of evaluating the proximity of a quantum state to the desired (ground state) by minimising the (overall) energy, aimed to minimise the energy of the lowest tail of a quantum state. This, intuitively, would succeed quicker in finding a quantum state that has a non-negligible overlap with the solution (but not necessarily very high overlap). This state, however, suffices to solve the problem. This intuition was also confirmed with numerical simulations. In other words, the cost function used in the classical optimiser, in order to find the optimal parameters, was not the energy of the quantum state, but the tail of the corresponding distribution. 

Inspired by this idea but also by adiabatic quantum computing \cite{albash2018adiabatic}, we consider here an evolving cost function. In our proposal the way that the cost of a quantum state is computed, dynamically changes during the classical optimisation process. We start with a cost function as in \cite{barkoutsos2020improving} focusing on a small tail, but during the optimisation process we gradually increase the tail (fraction of the distribution we ``count'') until we reach a point that all the distribution is included i.e. we measure the full expectation value of the energy  (as in ``standard'' cost functions).

\noindent\emph{Our contributions.} 
\begin{itemize}
    \item We introduce an evolving objective function that starts with the CVaR defined in \cite{barkoutsos2020improving} and gradually in the optimisation process becomes the full energy of the quantum state. Alternative forms of this Ascending-CVaR objective functions are considered and a linear and a sigmoid functions (that appear to perform better) are selected.
    \item We test our proposal, with classical  numerical simulations (using up to 20 qubits), both in the setting of VQE with hardware efficient ansatz and in QAOA. Our results suggest that our proposal leads to faster convergence with bigger overlap with the ideal solution than prior works, while crucially, succeeds in obtaining the solution in (many) instances that other techniques fail altogether (see Section \ref{conc} for statistics and comparisons).
    \item Our analysis is done for three different combinatorial optimisation problems, namely Max-Cut, Number Partitioning and Portfolio Optimisation. We consider many different instances and problem sizes where the conclusions persist in all cases. This has importance in its own right, since these problems are important by themselves, and our proposal gives an approach to improve the performance and bring closer achieving ``useful'' quantum advantage. Interestingly, our method offered greater advantage in ``hard instances'' of the problems, where the other methods frequently failed to find the solutions altogether.  
\end{itemize}

\noindent\emph{Structure.} In Section \ref{preliminaries} we give the essential background: We introduce the variational quantum algorithms and specifically, the variational quantum eigensolver and the quantum approximate optimisation algorithm. We then introduce the CVaR objective function of  
\cite{barkoutsos2020improving} and finally analyse the three different combinatorial optimisation problems that we use as case-studies. 
In Section \ref{ascending} we introduce our novel method called \emph{Ascending-CVaR} and we discuss the hyperparameters of our model. In section \ref{visualise} we illustrate our method using a small instance as an example. In Section \ref{methods} we discuss our methodology. In section \ref{results} we present the results of our method compared to existing objective functions. We conclude in Section \ref{conc} with a general discussion of our method and future work.

\section{Preliminaries}
\label{preliminaries}

We introduce two of the main Variational Quantum Algorithms \cite{cerezo2020variational}, the CVaR objective function \cite{barkoutsos2020improving} and the three types of combinatorial optimisation problems that we use. 

\begin{figure*}
    \centering
    \includegraphics[width=10cm,height=10cm,keepaspectratio]{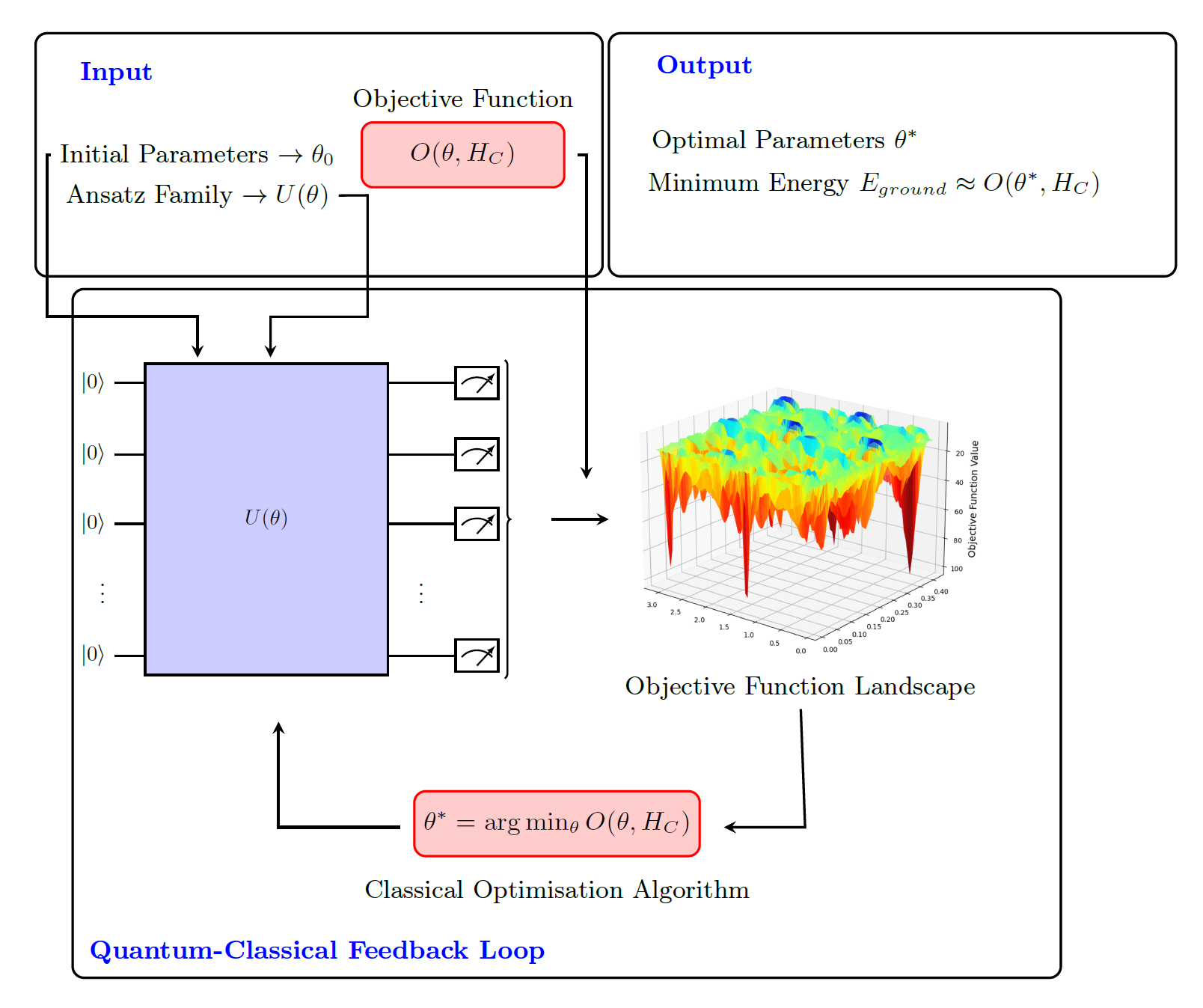} 
    \caption{General framework of a variational quantum algorithm. The optimisation problem, described by a cost function $C(\boldsymbol{x})$ is mapped to an interacting qubit Hamiltonian $H_C$. A parameterised family of states (``ansatz'') $U(\boldsymbol{\theta})$ with random initial parameters is chosen and a quantum-classical feedback loop iteratively updates the parameters $\boldsymbol{\theta}$. The optimisation ends when the stopping condition is met, and the optimal parameters $\boldsymbol{\theta^*}$ are outputted.}
    \label{fig:Variational_algorithm_framework}
\end{figure*}

\subsection{Variational Quantum Algorithms}

Here we revise the methods with focus on optimisation problems. 
The general framework of a variational quantum algorithm  is outlined in Figure \ref{fig:Variational_algorithm_framework}. The first step is to map the classical cost function $C(\boldsymbol{x})$ that describes the optimisation problem, into an interacting qubit Hamiltonian $H_C$ whose ground state gives the solution we are seeking.

The second step is to choose an ansatz family of unitary operators $U(\boldsymbol{\theta})$. This family is both efficiently expressible and trainable, parameterised by a $\mu$-parameter vector $\boldsymbol{\theta} = (\theta_1, ..., \theta_\mu)$ where $\mu = \mathcal{O}(poly(n))$ and $n$ is the system size. In general, the parameters are initialized at random\footnote{There are cases that a ``clever'' initialization could lead to faster convergence \cite{Sack2021quantumannealing, brandao2018fixed}}. The third step is to evaluate some objective function, usually taken to be the expectation value of the problem's Hamiltonian on the state considered $\bra{\psi(\boldsymbol{\theta})}H_C\ket{\psi(\boldsymbol{\theta})}$. This is done by preparing the state (applying the unitary $U(\boldsymbol{\theta})$ on the initial state) and then measuring the output in the computational basis and repeating this procedure for a given number of times (typically called ``shots''). This number determines the accuracy the objective function is evaluated. The fourth step is to update the parameters and repeat step three, iteratively using some classical optimiser until a stopping condition is satisfied.
We then say that the parameters are optimal, i.e.
\begin{equation}
   \boldsymbol{\theta ^*} = \arg\min_{\boldsymbol{\theta}} O \left(\boldsymbol{\theta}, H_C\right)
\end{equation}

The state produced by these parameters, $\ket{\psi(\boldsymbol{\theta^*})}=U(\boldsymbol{\theta^*})\ket{0}^{\otimes n}$, can be used to give an estimate of the ground state energy of the Hamiltonian $H_C$ and thus an approximate solution to the desired optimisation problem. The objective function used during this process, as stated above, typically coincides with the expectation value of the problem's Hamiltonian. However, we note here that other choices may also be possible, especially if we realise that the true target of the optimisation algorithm is to sample, at least once, the optimal solution. This can efficiently be produced if the output state has a sufficiently large (or more precisely simply non-vanishing) overlap with the optimal solution $\vert\bra{\psi(\boldsymbol{\theta^*})}\ket{\psi_{opt}}\vert^2$.

\subsection{Conditional Value-At-Risk}

Barkoutsos et al. \cite{barkoutsos2020improving} used an alternative objective function. They demonstrated that their proposal performed better than minimising the expectation value. The key observation is that for optimisation problems the optimal solution is a computational basis state. For computational basis states, one can compute their energy (efficiently). For a general quantum state $\ket{\psi(\boldsymbol{\theta})}$ one can prepare and measure it (multiple times) in the computational basis, and the expectation value of the energy is simply the average of the individual computational basis states energies. To find the overlap of this state with the optimal solution (ground state) one can simply observe the frequency of the computational basis state with the smallest energy. Naturally, if that overlap is too small (or even zero), it is possible that none of the measurements outcomes will give the solution. On the other hand it is also clear that the overlap of this state with computational basis vectors with high energy are irrelevant for finding the ground state. The idea of \cite{barkoutsos2020improving} was to use this observation and instead of using all the measurement outcomes and compute the expectation value, they used as objective function the lower tail of the distribution of energies obtained, i.e. ignored all but a small fraction (with smallest energy) of their measurement outcomes. 

They then demonstrated that their technique succeeded in getting quicker a quantum state that has a sufficiently large overlap with the ground state. This in turn, is sufficient to actually find this ground state, since as a final step, once the optimal $\boldsymbol{\theta^*}$ is found, one can keep the computational vector that has the smallest energy only. Specifically, let $H_k$ be the energy corresponding to a computational basis vector, and let us order them in such a way that larger $k$ corresponds to larger energy. For each state, one repeats the measurement $K$-times, so there are (up to) $K$ distinct values $H_k$. In \cite{barkoutsos2020improving} a new parameter $\alpha$ was introduced. Let $\alpha\in(0,1]$ be the fraction (part of the tail) that we want to keep.  This fraction, typically, needs to be non-negligible (we can assume for simplicity, that is constant). Then the objective function that was used, was the average of the smallest $\alpha K$ samples, i.e.

\begin{equation}
\label{eq:cvar}
    CVaR_\alpha = \frac{1}{\lceil{\alpha K}\rceil}\sum_{k=0}^{\lceil{\alpha K}\rceil}H_k
\end{equation}
In order to achieve the same accuracy when evaluating this objective function, as the accuracy achieved when computing the expectation value using $K$ shots, it is clear that the number of runs of the preparation circuit need to be increased to $K/\alpha$.

As it was proven by \cite{barkoutsos2020improving}, the angles $\boldsymbol{\theta^*}$ that minimise CVaR$_\alpha$ do not (in general) correspond to minima of the expectation value. As a result, the angles that lead to the smallest possible $\alpha$-tail differ from the angles that minimise the average of the samples. This fact motivates to introduce a lower $\alpha$-tail optimisation so as to achieve an overlap with the optimal state of at least $\alpha$, i.e find optimal $\boldsymbol{\theta^*}$ that satisfy:

\begin{equation}
|\bra{\psi(\boldsymbol{\theta^*})}\ket{\psi_{opt}}|^2 \geq \alpha
\end{equation}

\subsection{Combinatorial Optimisation Problems}

We test our proposed method in various instances of three different combinatorial optimisation problems. These are all important problems in their own right, so improving the performance of variational quantum algorithms for these problems is of independent interest. Moreover, testing our proposed objective function on different types of combinatorial optimisation problems demonstrates that improvements observed are generic and motivates further use for different applications. Given that our proposal's starting  point is the work of \cite{barkoutsos2020improving}, we included the problems that they tested their proposal to allow for more direct comparison.

The easiest way to use variational quantum algorithms for an optimisation problem is to first map the problem to a \emph{Quadratic Unconstrained Binary Optimization} (QUBO) problem. This is what we will do for all our examples. QUBO problems seek to solve (find the $\boldsymbol{x}$ that minimises the expression):

\begin{equation}
    \min_{\boldsymbol{x}} \left(b^{T}\boldsymbol{x} + \boldsymbol{x}^TA\boldsymbol{x}\right)
\end{equation}
where $b\in \mathbb{R}^n$ and $A \in \mathbb{R}^{n \times n}$. These cost functions can easily be mapped to an Ising Hamiltonian \cite{lucas2014ising} by first transforming the binary variables $x_i\in\{0,1\}$ according to:
\begin{equation}
x_i = \frac{1-z_i}{2}
\label{eq:transform}
\end{equation}
where $z_i\in\{-1,+1\}$ are spin variables, and then turning the cost function to a Hamiltonian by promoting these variables to
Pauli $\sigma^z_i$ operators, one for each qubit $i$. The QUBO problem then transforms to
\begin{equation}
\min_z c^{T}\boldsymbol{z} + \boldsymbol{z}^TQ\boldsymbol{z}
\label{eq:qubo}
\end{equation}
where the new $c\in \mathbb{R}^n$ and $Q \in \mathbb{R}^{n\times n}$ are easily computable.

Then, by replacing the spin variable $z_i$ with the Pauli $\sigma_i^z$ operator with corresponding eigenvalues $\{-1, +1\}$, the problem translates into finding the ground state, i.e. the spin configuration, of an $n$-qubit system interacting with the Hamiltonian:
\begin{equation}
H = \sum_{i=1}^n c_i \sigma_i^z + \sum_{i=1}^n Q_{ij} \sigma_i^z \sigma_j^z
\end{equation}

\subsubsection{Max-Cut Problem}

 The first problem is \emph{Max-Cut}. It is one of the most studied combinatorial problems in the context of variational quantum algorithms due to the simplicity and guaranteed performance at least for some instances
 \cite{farhi2014quantum, wang2018quantum}. 
 
 Let $G(V,E)$ be a non-directed $n$-vertex graph, where $V$ is the set of vertices, $E$ is the set of edges, and $w_{ij}$ are the weights of the edges. A \emph{cut} is defined as a bipartion of the set $V$ into two disjoint subsets $P,Q$, i.e. $P\cup Q = V$ and $ P\cap Q=\emptyset$. Equivalently, we label every vertex with either $0$ or $1$, where it is understood that the vertex belongs to set $P$ if it takes the value $0$ and to set $Q$ if it takes the value $1$. The aim is to maximise the following cost function:
 \begin{equation}
 C(\boldsymbol{x}) = \sum_{i,j = 1}^n w_{ij}x_i\left(1-x_j\right)
 \end{equation}
 This intuitively corresponds to finding a partition of the vertices into two disjoint sets that ``cuts'' the maximum number of edges. By applying the transformation, Eq. \eqref{eq:transform}, the cost function transforms into:
 \begin{equation}
     C(\boldsymbol{z}) = \sum_{\left<i,j\right>\in E} \frac{w_{ij}}{2}\left(1-z_iz_j\right)
 \end{equation}
Maximising the cost function above corresponds into finding the ground state of the Hamiltonian\footnote{Note the overall minus sign that turns the maximisation of the cost function to finding the minimum energy for the Hamiltonian.}:

\begin{equation}
	H_{C} = -\sum_{\left<i,j\right>\in E}\frac{w_{ij}}{2}\left( 1- \sigma_{i}^z \sigma_j^z \right)
\end{equation}

Max-Cut is known to be NP-Hard. The best classical approximation algorithm is that of Goemans and Williamson which uses semi-definite programming to achieve an approximation ratio, Eq. \eqref{eq:approx_ratio}, $r^* \approx 0.87856$ for all graphs. Note, that being NP-Hard implies that we do not expect to have an efficient quantum algorithm (poly-time) to solve the problem for its hardest instances\footnote{NP is strongly believed to not be included in BQP}, but we could definitely get improvements using quantum algorithms (either by smaller speed-ups or by heuristics that could solve more instances than classical heuristics).

Although it was proven that constant-depth QAOA does not outperform GW for certain class of problems \cite{bravyi2019obstacles}, there are instances where the approximation ratio of the former is larger than the latter \cite{crooks2018performance}. Note here that QAOA beats random guessing even at $p=1$ \cite{farhi2014quantum}, while Machine Learning techniques have been used to classify for which graph types is better to use QAOA instead of GW \cite{moussa2020quantum}. In general, however, the performance of QAOA in intermediate depths is still highly unexplored.

\subsubsection{Number Partitioning}

The second problem is \emph{Number Partitioning} and is stated as follows. Given a set of $N$ positive integers $S = \{n_1, n_2, ..., n_N\}$, the target is to find a bipartion of the set $S$ into two disjoint subsets $P,Q$, where $P \cup Q = S$ and $P\cap Q=\emptyset$ so that the difference between the sum of the elements on the set $P$ and the set $Q$ is minimized. We thus want to minimize the cost function:
\begin{equation}
C(\boldsymbol{x}) = \left(\sum_{i=1}^N(2x_i-1)n_i\right)^2
\end{equation}

The binary string $\boldsymbol{x}=x_1x_2\dots x_n$ corresponds to one configuration where a number $n_i$ is placed in the $P$ set ($x_i=0$) or in the $Q$ set ($x_i = 1)$.
The cost function can easily be mapped to the Ising Hamiltonian:
\begin{equation}
\label{eq:num_part_hamiltonian}
H_C = \left(\sum_{i=1}^N\sigma_i^zn_i\right)^2
\end{equation}

By expanding the cost function  Eq. \eqref{eq:num_part_hamiltonian}, the cost function can be written as:
\begin{equation}
    H_C = \sum_{i\neq j}(n_in_j)\sigma_i^z\sigma_j^z + \sum_{i=1}^Nn_i^2
\end{equation}

If we neglect the constant term, we can see that the Number Partitioning problem can be easily mapped to the \emph{Sherrington-Kirkpatrick model} which is an energy minimization problem with an all-to-all random couplings which was recently analysed on \cite{farhi2019quantum}.

Although the problem is known to be NP-Hard, it is also known as the ``easiest hard problem''. That is, because there exists a ``hard-easy'' phase transition \cite{mertens2006number} where instances belonging in the easy-phase can be efficiently tackled using heuristics \cite{korf2009multi}. Interestingly, it appears that one may be able to tackle some of the instances in the ``hard phase'' using variational quantum algorithms.

\subsubsection{Portfolio Optimisation}

The third problem is \emph{Portfolio Optimisation} \cite{orus2019quantum, venturelli2019reverse} and is stated as follows. Given a set of $n$ assets $\{0,\cdots,n\}$, corresponding expected returns $\mu_i$ and covariances $\Sigma_{ij}$, a risk factor $q > 0$ and a budget $B \in \{1,\ldots,n\}$, the considered portfolio optimisation problem tries to find a subset of assets $P\subset \{1,\ldots,n\}$ with $|P|=B$ such that the resulting \emph{q-weighted-mean-variance}, i.e. $\sum_{i\in P} \mu_i - q\sum_{i,j\in P}\Sigma_{ij}$, is maximised. In other words, we want to maximise the cost function:

\begin{equation}
C(\boldsymbol{x}) = \sum_{i=1}^n \mu_i x_i - q\sum_{i,j=1}^n\Sigma_{ij}x_ix_j
\label{eq:portopt}
\end{equation}
along with the constraint
\begin{equation}
    \sum_{i=1}^nx_i = B
\label{eq:constraint}
\end{equation}

The \emph{portfolio vector} $x\in\{0,1\}^n$, consisting of $n$ binary decision variables, indicates whether an asset is picked ($x_i = 1$) or not ($x_i = 0$). The constraint in (\ref{eq:constraint}) is translated as an extra penalty term in the Hamiltonian $(\sum_{i=1}^nx_i-B)^2$.

The problem is known to be NP-complete \cite{kellerer2000selecting}. We apply the transformation, Eq. \eqref{eq:transform}, so the cost function transforms into:
\begin{equation}
\begin{aligned}
C(\boldsymbol{z}) &= - q\sum_{i,j=1}^n \frac{\Sigma_{ij}}{4}z_i z_j + \sum_{i=1}^n\left(\sum_{j=1}^n \frac{q\Sigma_{ij}z_i}{2}-\frac{\mu_iz_i}{2}\right) \\&+\sum_{i=1}^n\left(\frac{\mu_i}{2} - \sum_{j=1}^n\frac{q\Sigma_{ij}}{4}\right)
\end{aligned}
\end{equation}
which, along with the extra penalty term, corresponds to minimising the Hamiltonian :
 \begin{equation}
 \begin{aligned}
     H_C &=  \sum_{i,j=1}^n \frac{q\Sigma_{ij}}{4}\sigma_i^z\sigma_j^z - \sum_{i=1}^n\left(\sum_{j=1}^n \frac{q\Sigma_{ij}\sigma_i^z}{2}-\frac{\mu_i\sigma_i^z}{2}\right) \\
     & -\sum_{i=1}^n\left(\frac{\mu_i}{2} - \sum_{j=1}^n\frac{q\Sigma_{ij}}{4}\right) + \left(\sum_{i=1}^n\sigma_i^z + \frac{n}{2}-B\right)^2
 \end{aligned}
 \end{equation}

Portfolio optimisation as given in Eq. (\ref{eq:portopt}) was recently tackled using variational quantum algorithms \cite{egger2020quantum}, using warm-starting QAOA \cite{egger2020warm} and on D-wave systems using quantum annealing \cite{cohen2020portfolio}. Prior to our work, \cite{Slate2021quantumwalkbased} developed a quantum-walk-based optimisation algorithm and \cite{mugel2020dynamic} considered a more general setting of portfolio optimisation, called dynamic portfolio optimisation, where one has to allocate weights to a number of assets in a period of time in order to maximise the overall return.

\section{Ascending-CVaR}
\label{ascending}

The CVaR cost function of \cite{barkoutsos2020improving} was shown to perform better in general, than the ``standard'' expectation value. There are three observations, however, that motivates our proposal. First, as noted in \cite{barkoutsos2020improving}, the choice of $\alpha$ is somehow random, and importantly, for different problems and even for different instances of the same class of problems, the optimal choice of $\alpha$ varies in a non-obvious (e.g. monotonic) way. The performance of the algorithm's speed, but also if it finds the solution at all, depends on that choice. The second point is that optimising with a fixed small $\alpha$ has further disadvantages: (i) it ``finds'' parameters $\boldsymbol{\theta}$ that result to a state that does not have the greatest overlap with the solution and (ii) the true running time of the algorithm to achieve same accuracy is larger, in other words for each iteration one requires $1/\alpha$ times more measurements to achieve the same accuracy in estimating the cost function (since only the lower $\alpha$ fraction of the measurements are used). Finally, the third observation is that the $CVaR_{\alpha}$ objective functions with different $\alpha$ have a different energy landscape. For any fixed choice of $\alpha$ the optimiser could ``get stuck'' at a local minimum. Interestingly, if one varies the $\alpha$ during the optimisation, while we still ensure that if the algorithm finds the true ground state it remains there, we also avoid getting stuck at local minima since those are different for different choices of $\alpha$. Therefore if the optimiser reaches a point that has a local minimum for one value of $\alpha$, when $\alpha$ changes this point (may) no longer be a local minimum and thus could continue ``moving'' towards the true global minimum (ground state).

Let's say that an optimisation problem has an optimal solution which is a computational basis state and we denote it as $\ket{\psi_{opt}}$. Let's also assume that a parameterised family of gates, $U(\boldsymbol{\theta})$, acts on the $\ket{0}^{\otimes n}$ state and produces the state
\begin{equation}
\label{eq:general_state}
\ket{\psi(\boldsymbol{\theta})} =  a_{opt}(\boldsymbol{\theta})\ket{\psi_{opt}} + (1-a_{opt}(\boldsymbol{\theta}))\ket{\psi_{other}}
\end{equation}
where $\ket{\psi_{other}}$ is the superposition of all sub-optimal computational basis states. Let's also assume that this parameterised family of states can achieve \emph{a maximum overlap $\kappa$ with the optimal solution}\footnote{In other words, the complex coefficient $a_{opt}(\boldsymbol{\theta})$ corresponding to the probability of sampling the optimal solution $Prob(opt) = |a_{opt}|^2$ has a maximum value : $\max_{\boldsymbol{\theta}} |a_{opt}(\boldsymbol{\theta})|^2 = \kappa$)}. We can write the state $\ket{\psi}$, corresponding to the state with the highest overlap, without loss of generality as:

\begin{equation}
\ket{\psi} = \sqrt{\kappa} \ket{\psi_{opt}} + \left(1-\sqrt{\kappa}\right) \ket{\psi_{other}} 
\end{equation}

\noindent\textbf{Proposition 1.} \emph{For the family of states in Eq. \eqref{eq:general_state} and for all $\alpha \leq \kappa$:
\begin{equation}
\min_{\boldsymbol{\theta}} CVaR_{\alpha}(\boldsymbol{\theta}) = \min_{\ket{\phi}} \bra{\phi}H_C\ket{\phi}
\end{equation}
i.e. all $CVaR_{\alpha}$ with $\alpha \leq \kappa$ share the same minimum objective function value which is the smallest eigenvalue of the Hamiltonian.}
\vspace{2mm}

It is clear from Proposition 1 that \emph{all $CVaR_{\alpha}(\boldsymbol{\theta})$ with $\alpha \leq \kappa$ share the same ground state}, which is the true optimum of the optimisation problem. Thus all angles $\boldsymbol{\theta^*}$ that correspond to a global minimum of $CVaR_{\alpha_1}$ will also correspond to a global minimum of $CVaR_{\alpha_2}$ if $\alpha_2 \leq \alpha_1 \leq \kappa$. For example for an ansatz $U(\theta)$ that is able to attain 10\% overlap with the optimal computational basis state, if one is able to find the global minimum of $CVaR_{0.1}$, which means that 10\% of the measurements correspond to the ground state, then it is clear that all $\alpha < 0.1$ will also be minimised by the same angles.
\vspace{2mm}

\noindent\textbf{Proposition 2.} \emph{Let an optimisation problem with an optimal solution $\ket{\psi_{opt}}$ corresponding to a computational basis state. For any parameterised family of gates $U(\boldsymbol{\theta})$ that can achieve a maximum overlap $\kappa$ with the optimal solution, the angles $\theta^*$ that correspond to the global minimum of $CVaR_{\alpha_1}$ will also correspond to a global minimum for $CVaR_{\alpha_2}$ if $\alpha_1 \leq \alpha_2 \leq \kappa$. The converse does not necessarily hold.}
\vspace{2mm}

In other words, Proposition 2 states that if $\boldsymbol{\theta^*} = \arg\min_{\boldsymbol{\theta}}CVaR_{\alpha_2}(\boldsymbol{\theta})$ then also $\boldsymbol{\theta^*} = \arg\min_{\boldsymbol{\theta}}CVaR_{\alpha_1}(\boldsymbol{\theta})$ for all $\alpha_1 \leq \alpha_2 \leq \kappa$. This indicates why decreasing $\alpha$ may not seem like a good choice. If for example the optimiser is able to find the optimal angles that minimise $CVaR_{\alpha_1}$ with $\alpha_1\leq\kappa$, then for all $\alpha_2<\alpha_1$ they will still remain optimal angles and thus will not be able to achieve a higher overlap state.
\vspace{2mm}

\noindent\textbf{Proposition 3.} \emph{A local minimum for $CVaR_{\alpha_1}$ does not necessarily correspond to a local minimum for $CVaR_{\alpha_2}$ if $\alpha_1 \neq \alpha_2$.}
\vspace{2mm}

Proposition 3 was proven using a counterexample in \cite{barkoutsos2020improving}. All these Propositions are important for introducing a non-stationary optimisation technique that avoids local minima. We know from Proposition 1 that all $CVaR_{\alpha}$ objective functions with $\alpha \in (0, \kappa]$ share the same minimum objective value which is the ground state energy of the Hamiltonian. We also know from Proposition 2 that many of the global minima for $\alpha_1$ may not be a global minimum for $\alpha_2$ if $\alpha_1 < \alpha_2$ and thus increasing $\alpha$ introduces extra information about the optimality of states. Finally, Proposition 3 indicates that different objective functions are associated with different energy landscapes as they do not agree on the local minima.

However, knowing the maximum overlap $\kappa$ in advance is not always possible. In the case of VQE with a hardware efficient ansatz, it can be shown that $\kappa =1$ and so $\min_{\boldsymbol{\theta}} CVaR_{\alpha}(\boldsymbol{\theta}) = \min_{\ket{\phi}} \bra{\phi}H_C\ket{\phi}$ for every $\alpha \in (0, 1]$. On the other hand for the QAOA ansatz, our experiments showed that $\kappa$ is usually small on low depth but increases with the number of layers

The cost functions used in variational quantum algorithms, to our knowledge, are ``constant in time'', meaning that the whole optimisation is run with a fixed cost function. To solve the issue of ``selecting the best $\alpha$'' and the other reasons listed above, we propose to use a dynamically evolving cost function, that essentially passes through a fixed set of $\alpha$ values. In the case of VQE, it is initialised in a very small value and the optimisation ends with $\alpha=1$ that is the standard expectation value of the Hamiltonian. We call all these cost functions \emph{Ascending-CVaR}. This has also a great(er) number of free choices, since we can now freely choose the (ascending) function. However, all choices we tried for the ascending function performed (in general) better than fixed $\alpha$, which indicates that the evolving cost function is a promising approach. For the remaining of the paper we focused on two functions that performed better:

The \emph{linear ascending} in which the parameter $\alpha_t$ is iteratively and discretely increased by the rule:

\begin{equation}
\begin{aligned}
\alpha_{t+1} &= \alpha_t + \lambda \\
CVaR_{\alpha_t} &= \frac{1}{\lceil{\alpha_t K}\rceil}\sum_{k=0}^{\lceil{\alpha_tK}\rceil}H_k
\label{eq:lin_ascending}
\end{aligned}
\end{equation}
where $\lambda \in [0.025, 0.045]$ is the \emph{ascending factor} and $0< \alpha_t \leq 1$.

The \emph{sigmoid ascending} in which the parameter $\alpha_t$ is discretely increased according to the function :

\begin{equation}
\label{eq:sig_ascending}
\begin{aligned}
\alpha_t &= \frac{1}{1+e^{5 - \lambda t}}
\end{aligned}
\end{equation}
where $\lambda \in [0.3, 0.4]$ is again the ascending factor and $0 < \alpha_t \leq 1$.

\begin{figure}
\begin{tikzpicture}
\node (img)  {\includegraphics[scale=0.32]{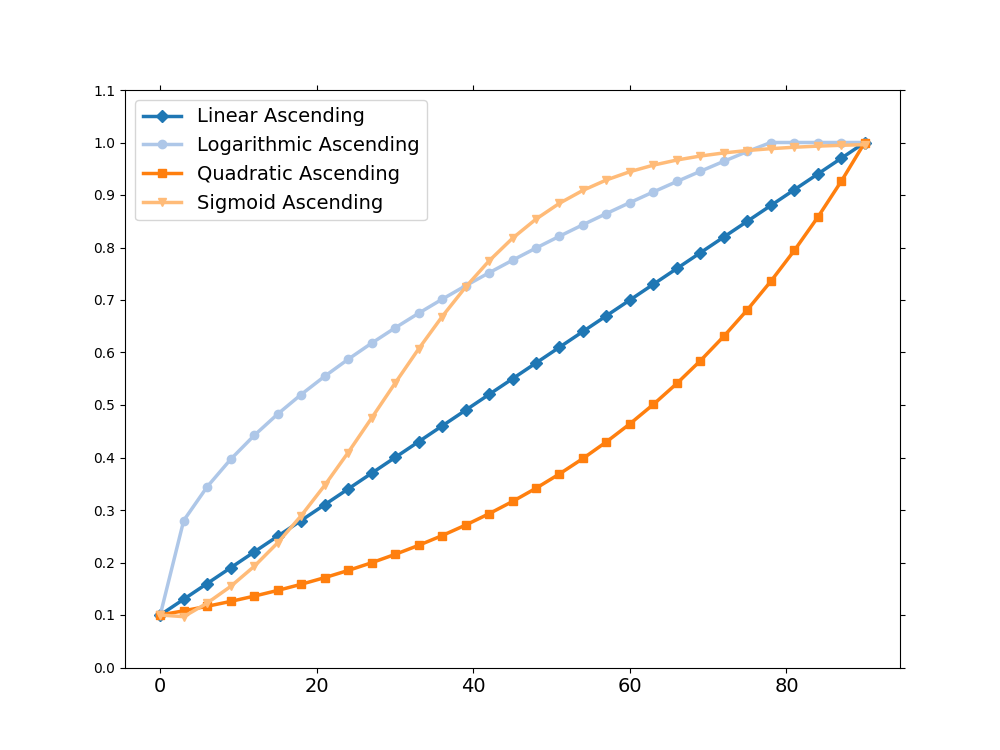}};
\node[below=of img, node distance=0cm, yshift=1.5cm] {\scriptsize Normalised Optimiser Iterations};
\node[left=of img, node distance=0cm, xshift=0.7cm, rotate=90, anchor=center,yshift=-0.7cm] {\scriptsize Prob. of Optimal Solution};
\end{tikzpicture}
\caption{Different choices for the ascending function. All functions start from the same initial point, $\alpha_0=0.01$ and ascend until $\alpha_f=1$ is reached.}
\label{fig:different_functions}
\end{figure}

To reach this conclusion we tested four different functions, a \emph{sigmoid}, a \emph{linear}, an \emph{exponential} and a \emph{logarithmic} (see Figure \ref{fig:different_functions}) on VQE-CVaR${_{\alpha_t}}$ with various different ascending rates. All functions were tested on all three problems. The metrics used were the magnitude of the overlap with the optimal solution, the success rate (i.e. the number of times where it succeeds to achieve a non-negligible overlap) as well as the average time taken to achieve at least $10\%$ overlap (for details see \ref{methods}).

\begin{figure}
\begin{tikzpicture}
\node (img)  {\includegraphics[scale=0.32]{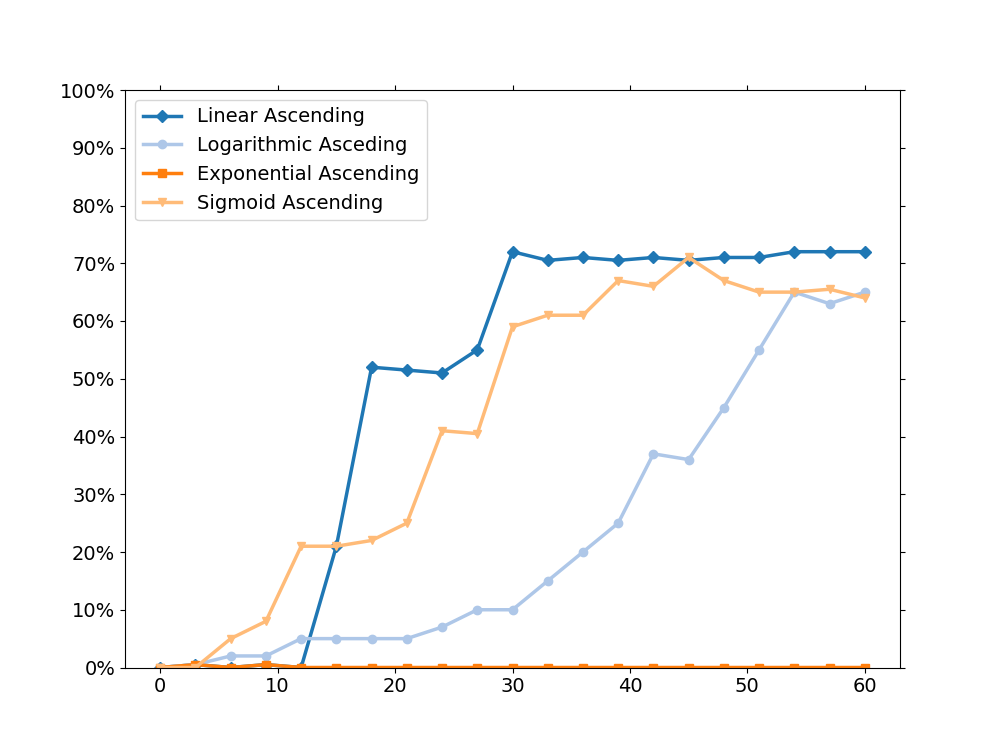}};
\node[below=of img, node distance=0cm, yshift=1.5cm] {\scriptsize Normalised Optimiser Iterations};
\node[left=of img, node distance=0cm, xshift=0.7cm, rotate=90, anchor=center,yshift=-0.7cm] {\scriptsize Prob. of Optimal Solution};
\end{tikzpicture}
\caption{Portfolio optimisation instance for 18 assets and different ascending functions. The blue line (down-pointing marker) indicates the linear ascending and always achieves a high overlap with the optimal solution in contrast to the orange line (line marker), the exponential ascending, which fails in almost any instance.}
\label{fig:different_functions_portfolio}
\end{figure}

The linear ascending, Eq. \eqref{eq:lin_ascending}, and the sigmoid ascending, Eq. \eqref{eq:sig_ascending}, functions have the most steady behavior as it can be seen in Figure \ref{fig:different_functions_portfolio}, outperforming the other two types on the majority of instances. The sigmoid was slightly slower in terms of speed, which is why we mainly used the linear one. However, as we will discuss in the next section, it appears that it may be better in some classes of problems especially on harder instances with $\sim 50$ qubits. It seems that in those cases, the optimiser is doing better spending the majority of its iterations on low $\alpha$ values and thus the sigmoid performs better. 

It is worth noting that increasing $\alpha$ to $\alpha = 1$ where it becomes the expectation value is not necessary. The whole point of variational algorithms is to achieve a constant non-negligible overlap with the optimal or near-optimal solution. For that reason one could only vary $\alpha$ until it reaches a threshold $\epsilon$ truncating the optimisation and reducing the number of iterations by a considerable amount.

We remark here that \emph{Ascending-CVaR} is fundamentally different from an adaptive strategy that selects the optimal value for the parameter $\alpha$. Specifically, by looking at the results in Section \ref{results} we can see that our method is able to reach quantum states that result in a very high overlap with the optimal state (almost equal to unity) that no constant choice of $\alpha$ would be able to attain. Even when the question is whether we find the solution with at least some small probability, Ascending-CVaR succeeds in cases where all of the fixed $\alpha$ failed.

In order to get the intuition why our method works one should think of what varying $\alpha$ actually does. The optimiser, at each iteration, moves towards a local (or a global) minimum corresponding to the instantaneous value of $\alpha$. By increasing $\alpha$ at every step, the optimisation is able to ``see'' a larger part of the energy distribution of the quantum state. This translates to the optimiser gaining additional information about the quantum state which modifies the objective function landscape.

This extra information alters the landscape and is thus able to ``erase'' false local minima, while using the previous step as ``initialisation'' it is unlikely that the optimiser will get stuck in new sub-optimal minima. Moreover, since only the global minima are invariant under $\alpha$ transformations, this change in the landscape will not affect any correct moves of the optimiser. In other words, if the optimisation algorithm did converge in a sub-optimal local minimum for a value of $\alpha$, it may not still be in a local minimum by switching into a different value of $\alpha$. Hence, the \emph{Ascending-CVaR} algorithm is guiding the trajectory of the optimiser in the highly parameterised space until it reaches a (nearly) optimal solution.

Finally, the reason our method is sensitive to the choice of the ascending factor $\lambda$ is related to the speed that the optimiser is receiving this extra information (see Appendix \ref{appendix:numerical_analysis} on how we numerically test the ascending factors for the different optimisation problems).

The pseudocode for the Ascending-CVaR algorithm is outlined in Algorithm \ref{alg:Asc-CVaR}

\begin{algorithm}
\caption{General Ascending-CVaR Optimisation Algorithm}
\label{alg:Asc-CVaR}
\SetKwInOut{Input}{Require}
\Input{Cost Function $C(\boldsymbol{\theta})$\;
 $\boldsymbol{\theta^{(0)}} \leftarrow$ Random initial parameters in the domain of  $C(\boldsymbol{\theta})$\;
 $\alpha_0 \leftarrow$ Initial $\alpha$\;
 $g(\alpha) \leftarrow$ Ascending function\;
 $U(\boldsymbol{\theta}) \leftarrow $ Ansatz Family
}
\For {$i=1,2,\dots$}{
    $\boldsymbol{\theta^*} = \arg\min_{\boldsymbol{\theta}}CVaR_{\alpha_{i-1}}(\boldsymbol{\theta})$ with initial parameters $\boldsymbol{\theta^{(0)}}$\;
    \If{stopping condition is met}{
        \Return $\boldsymbol{\theta^*}$\;
        }
    $\alpha_i \leftarrow g(\alpha_{i-1})$\;
    $\boldsymbol{\theta^{(0)}} = \boldsymbol{\theta^*}$\;
    }
\end{algorithm}

\section{Why our method works: An example}
\label{visualise}

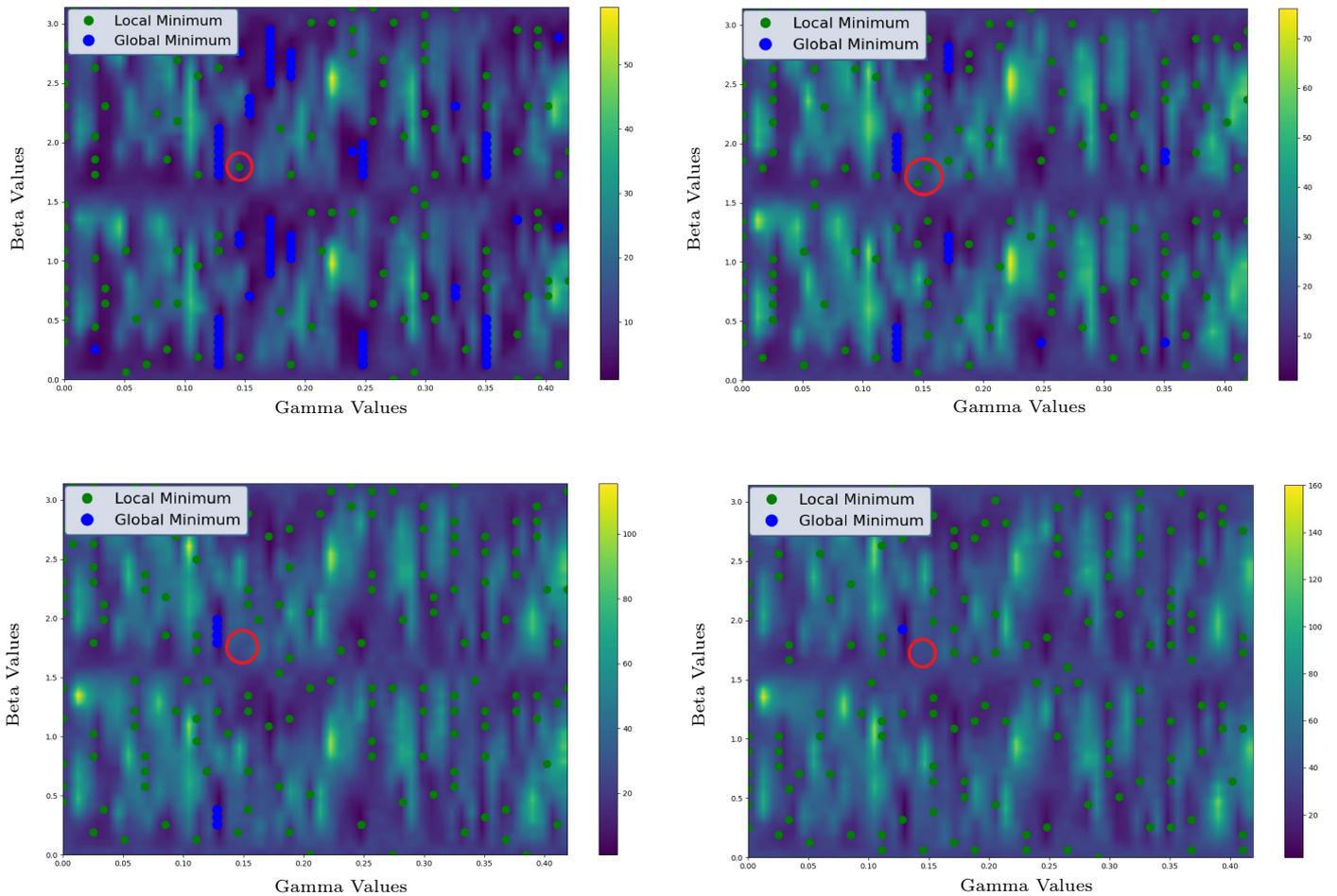
\begin{figure*}
\begin{tikzpicture}
\node (img1)  {\includegraphics[scale=0.285]{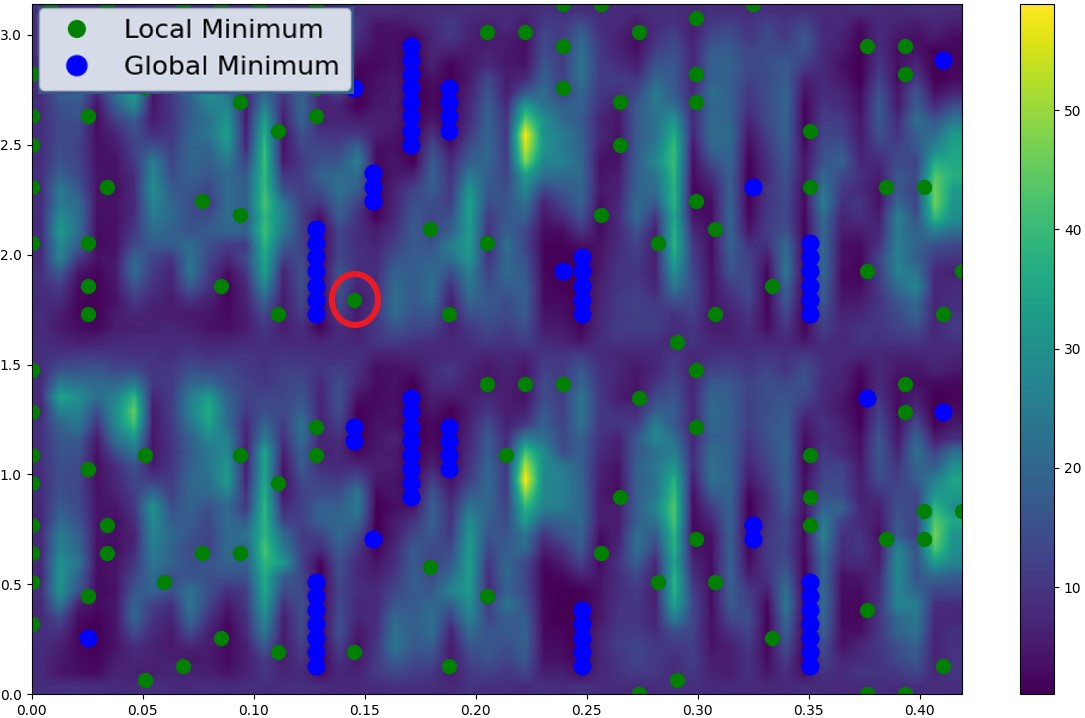}};
\node[below=of img1, node distance=0cm, yshift=1.1cm] {\scriptsize Gamma Values};
\node[left=of img1, node distance=0cm, rotate=90, anchor=center,yshift=-0.7cm] {\scriptsize Beta Values};
\node[right=of img1] (img2)  {\includegraphics[scale=0.285]{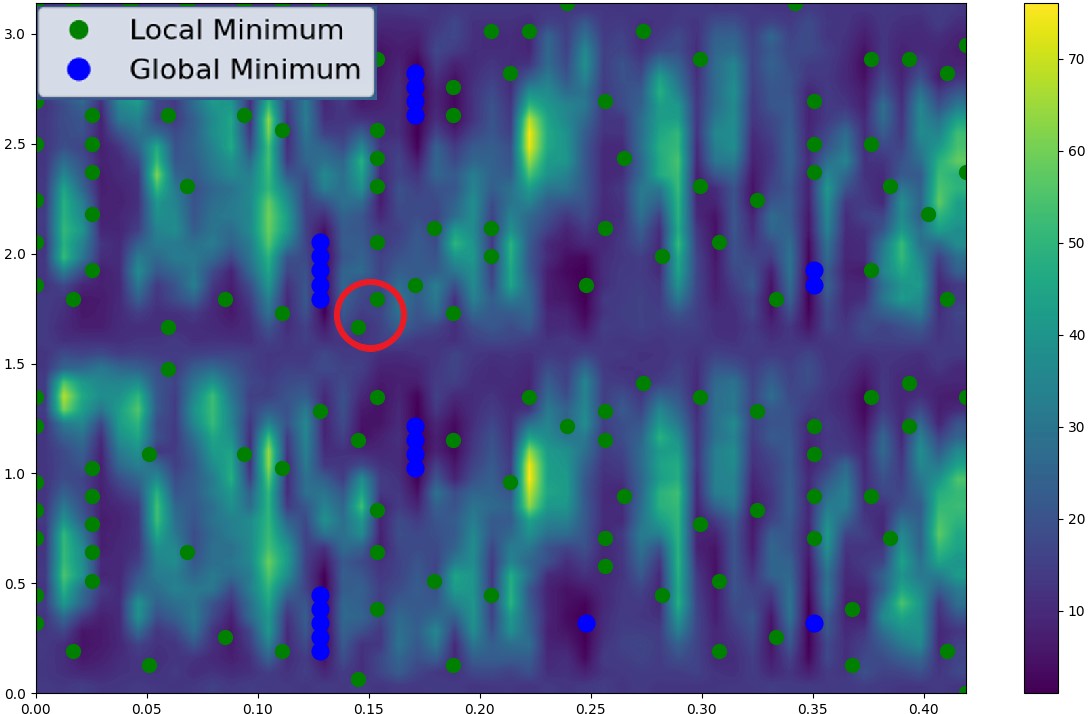}};
\node[below=of img2, node distance=0cm, yshift=1.1cm] {\scriptsize Gamma Values};
\node[left=of img2, node distance=0cm, rotate=90, anchor=center,yshift=-0.7cm] {\scriptsize Beta Values};
\node[below=of img1] (img3) {\includegraphics[scale=0.285]{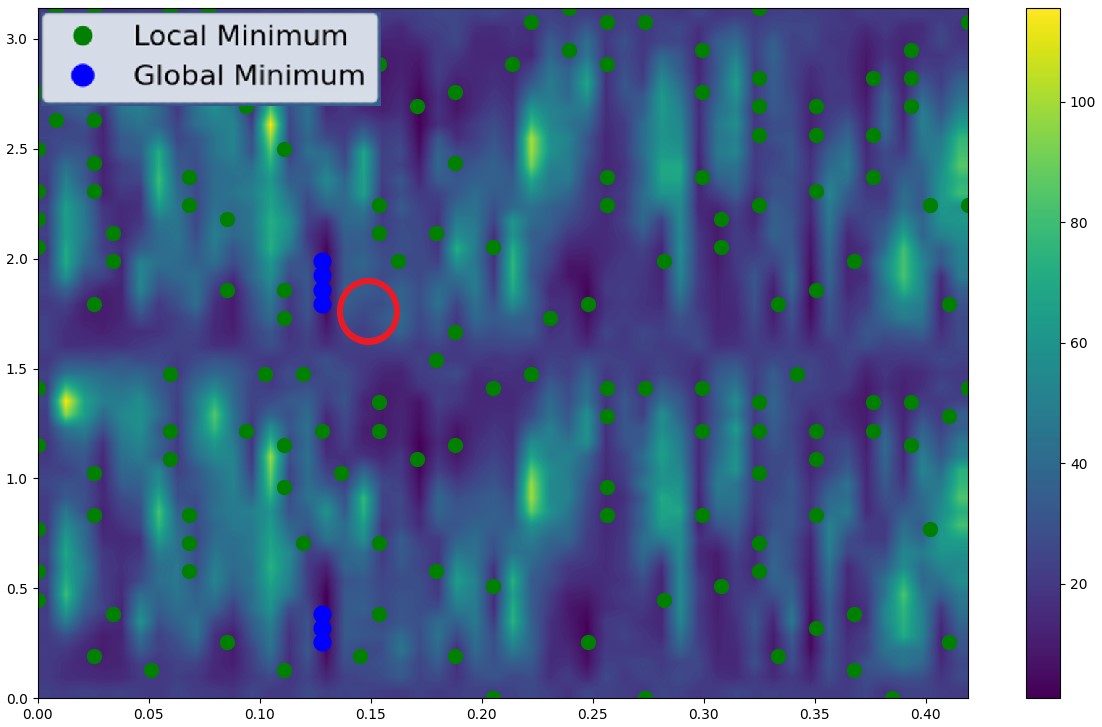}};
\node[below=of img3, node distance=0cm, yshift=1.1cm]{\scriptsize Gamma Values};
\node[left=of img3, node distance=0cm, rotate=90, anchor=center, yshift=-0.7cm] {\scriptsize Beta Values};
\node[right=of img3] (img4)  {\includegraphics[scale=0.285]{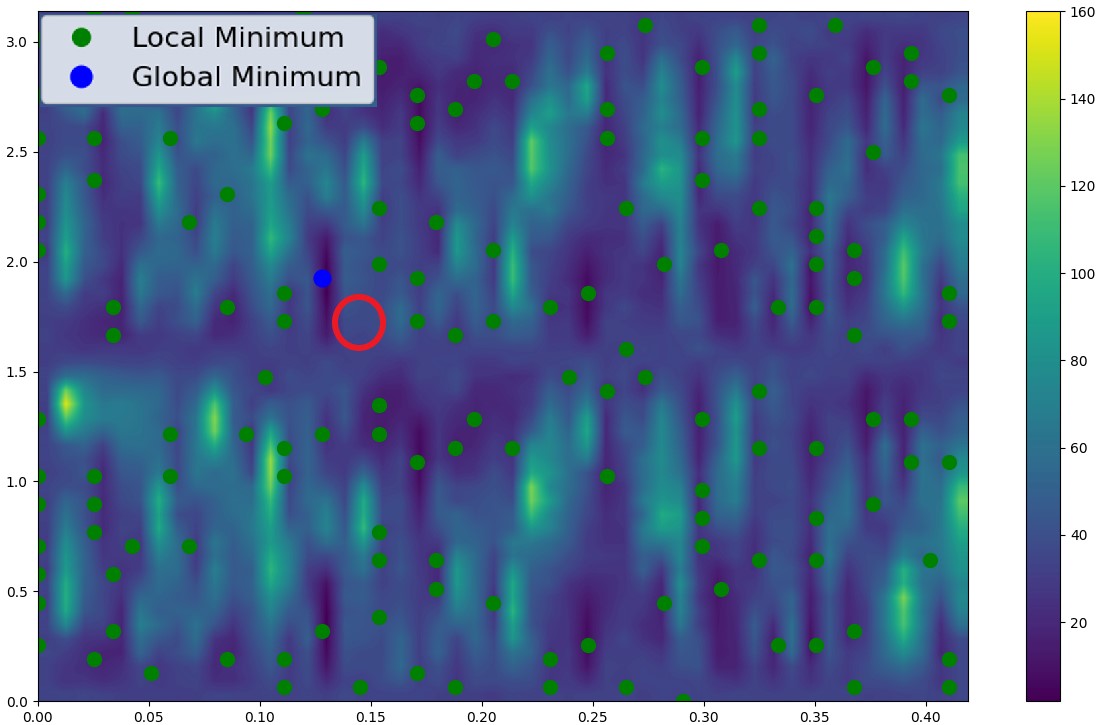}};
\node[below=of img4, node distance=0cm, yshift=1.1cm] {\scriptsize Gamma Values};
\node[left=of img4, node distance=0cm, rotate=90, anchor=center,yshift=-0.7cm] {\scriptsize Beta Values};
\end{tikzpicture}
\caption{Visualisation of local and global minima for different $CVaR_{\alpha}$ objective functions. On the top two figures, corresponding to $\alpha=0.05$ on the left and $\alpha=0.08$ on the right, you can see the local minima drawn in the red circle. However, on the bottom figures, corresponding to $\alpha=0.11$ on the left and $\alpha=0.14$ on the right, the local minima no longer exist.}
\label{fig:change_of_local_minima}
\end{figure*}

It would be illustrative to describe how local minima may vanish when the objective function is changed during the optimisation. In Figure \ref{fig:change_of_local_minima} we plot the $CVaR_{\alpha}$ objective function landscape for different values of $\alpha$. We choose to draw the landscape for the QAOA algorithm with depth $p=1$ because the two parameters $\beta, \gamma$ make it suitable to visualise in a 2D plot. On the contrary, VQE with a hardware efficient ansatz even on depth $p=1$ would require $2n$ parameters.

It can be easily seen that the positions of the local minima change but the position of the true global minimum remain the same while the condition $\alpha \leq \kappa$ holds  (see Proposition 1). However, in order to make it clearer for the reader, we choose to circle the position of a local minimum, located at $\gamma=0.15,\beta=1.75$. In this case, we can see how the local minima vanishes during the variation of the objective function. An optimiser that could stuck during the optimisation on a fixed value of $\alpha$ could ``unstuck'' with the change of $\alpha$.

The problem corresponding to the figures is a small instance of the Number Partitioning problem with size $n=8$. Even in a small size instance like this, the landscape is full of sub-optimal local minima where the optimiser could falsely converge to. This case-problem however does not constitute an example to prove the value of our method but only to visualise the changes in the energy landscape. The biggest improvements were observed in high dimensional expressive parameterised family of gates like the VQE with a hardware efficient ansatz or larger depth QAOA which cannot be plotted in a two-dimensional contour.

\section{Methods}
\label{methods}

One common metric used, especially in QAOA, is the approximation ratio as given in Eq. (\ref{eq:approx_ratio}). However, as we noted earlier, the true aim of variational quantum algorithms for combinatorial optimisation is to obtain quickly a sufficiently high (but not necessarily close to unity) overlap with the optimal solution. The CVaR method, for example, is constructed in a way that the maximum overlap achieved is not unity but determined by the risk $\alpha$. While our approach does achieve high approximation ratio, to make a fair and more complete comparison with prior works and importantly with \cite{barkoutsos2020improving}, we use different metrics. Specifically,
to benchmark and test our proposed method, we used three different types of metrics. The first is the overlap with the optimal solution. If $\ket{\psi_{opt,i}}$ is a $d$-degenerate ground state of the problem Hamiltonian, then the overlap is defined as:

\begin{equation}
\sum_{i=1}^d|\bra{\psi(\theta)}\ket{\psi_{opt,i}}|^2
\end{equation}
i.e. the probability of obtaining the optimal solution, given the parameters $\theta$. It follows that the parameterised state with the highest overlap with the optimal solution leads to sampling that optimal solution with the least number of circuit executions.

The second metric we want to test is the time taken to reach a given fixed overlap. We set a threshold of $10\%$ probability of obtaining the optimal solution and we tested which method achieves at least that probability faster. We note however that, in order to test which method converges to a $10\%$ overlap faster, we have to use $\alpha\geq 0.1$ because all $\alpha<0.1$ are not guaranteed to converge in an overlap of $10\%$ since the parameters $\theta$ than minimise $\alpha$ lead in an overlap smaller than 0.1.

To summarise the results and compare better the different approaches, for each cost function we divided the problem instances to those that the cost function is successful and to those that it fails. The meaning of what constitutes a ``successful'' run or a ``failed'' run cannot be unambiguously defined. For our work we consider that an optimiser is successful at a given instance of a problem if it achieves at least $10\%$ overlap with the optimal solution. It is clear that as the size of the problem instances increase, achieving a fixed $10\%$ overlap becomes harder\footnote{We should note that even a much smaller overlap is sufficient to find at least once the solution, provided that the number of ``shots'' is sufficiently large.}. In our analysis we chose $10\%$ since this leads to interesting behaviour where the methods analysed differ in their performance.

In our experiment, for comparing with fixed $\alpha$ we used four different choices: $\alpha=0.1, 0.2, 0.5, 1$. The $\alpha=1$ choice corresponds to a non CVaR objective function. Specifically, $\alpha = 1$ refers to the expectation value (it includes all the measurement outcomes) and it is the objective function that has been used in the overwhelming majority of the existing literature on variational quantum algorithms. In \cite{barkoutsos2020improving} they did an extensive comparison of CVaR with the expectation value (on the same combinatorial problems we make our analysis). For that reason, we choose to make the comparison of all different choices of $\alpha$ with our proposed $\alpha_t$ and plot our results in one section (see Section \ref{results}).

We also note that ascending factors $\lambda \in [0.025, 0.045]$ and $\lambda \in [0.3, 0.4]$ were found to be a good choice for the three different problems on instances with 15 to 20 qubits for the linear and sigmoid ascending respectively. (see Appendix \ref{appendix:numerical_analysis}). However, we would like to stress that if the sizes of the instances increase or even if the problems change but the sizes remain the same, one would have to readjust the hyperparameter $\lambda$. Investigating theoretically the choice of both the ascending factor and the ascending function given the characteristics of the problem as well as possible connections of our method to adiabatic quantum computing goes beyond the scope of this paper but will be investigated in a subsequent work.

In the QAOA algorithm we tested instances using depth $p=1$ to $p=6$ while on VQE we worked only on the depth $p=1$, since this depth was sufficient to get very good accuracy. In near term devices for the QAOA algorithm, increasing the depth even more becomes impractical due noise and decoherence.  
For this reason we did not consider greater depth, despite the fact that theoretically this could lead to better performance. This means that the variational ansatz for QAOA has only $2$ to $12$ parameters, i.e. only a fraction of the total parameters present in hardware efficient ansatz used for VQE in depth-1.

To account for the different sizes of problem instances, and to make a fair comparison for the speed of convergence, we used the normalised optimiser iterations \cite{more2009benchmarking}. Note that this choice is made in order to be able to compare the performance of the algorithm among instances that involve different number of qubits, and see how the improvement offered by Ascending-CVaR is independent of the instance size. Concretely, the normalised optimiser iterations is defined as the number of times the optimiser evaluates the objective function divided by the function's number of parameters, i.e. the number of parameters of the ansatz. In the case of the VQE the number of parameters are $n(1 + p)$ while on QAOA are $2p$. We note however, that the real time of convergence could be used as seen in Appendix \ref{appendix}, where we compare the performance with respect to the total number of circuit repetitions. However, as we show below, there are instances where the constant CVaR does not achieve even a small overlap with the optimal solution and in those cases the time taken becomes irrelevant.

 We ran our experiments on IBM's \emph{Qiskit Aer} simulator, allowing noiseless multi-shot executions of our circuit. We set the number of executions of our circuit to $K=1000$, which scaled up as $K/\alpha$ with the choice of $\alpha$. All instances were given a maximum of $\left(66\times parameters\right)$ optimiser iterations which is more than enough iterations for an optimiser to converge to a minimum in the problems we implemented. They were initialised with a random choice of parameters, but the same for all different choices of $\alpha$. We used the same gradient-free optimiser, COBYLA \cite{powell1994direct}, for all different problems and instances as it was shown to outperform other classical optimisers \cite{nannicini2019performance}. 

\section{Results}
\label{results}

We will analyse the results for each of the three combinatorial optimisation problems separately. For each of them we will present the results for VQE with hardware-efficient ansatz first and then the results for QAOA. We note that for all three combinatorial optimisation problems and for all methods used (Ascending-CVaR, constant CVaR, expectation value), VQE performs (much) better than QAOA, at least for the sufficiently shallow circuits that we consider. Our method improves the performance in both cases (VQE, QAOA) but since VQE gives much better results for these problems, in the comparison and discussion we will focus on VQE instances only.

\subsection{Max-Cut}

For the \emph{Max-Cut} problem we worked on unweighted graphs with 15-19 vertices, drawn from different graph classes and sampled them using the NetworkX library \cite{hagberg2008exploring}.

\subsubsection{\texorpdfstring{CVaR$_{\alpha_t}$-VQE}{Lg}}

\begin{table*}
\begin{center}
\begin{tabular}{||c||c|c|c|c|c||c|c|c|c|c||}
    \hline
    Max-Cut&\multicolumn{5}{c||}{Successful Instances}&\multicolumn{5}{c||}{Average Overlap}\\
    \hline
     & $\alpha_t$ & 0.1 & 0.2 & 0.5 & 1 & $\alpha_t$ & 0.1 & 0.2 & 0.5 & 1 \\
     \hline
     Random Graphs  & 96 & 84 & 81 & 68 & 53 & 64.69 & 12.13 & 21.45 & 39.28 & 36.24\\
     \hline
\end{tabular}
\caption{Results table for the \emph{Max-Cut} problem (VQE) for 100 random non-regular unweighted graph instances with 15 to 19 vertices.}
\label{table:max_cut_table}
\end{center}
\end{table*}

For regular graphs, CVaR$_{\alpha_t}$-VQE behaved equally well with constant-$\alpha$'s optimisation as well as with the expectation value. All of the methods reached the chosen threshold of $10\%$ overlap with the optimal state at almost equal times without any difficulty. For that reason we focused on harder non-regular instances where our method outperformed the latter methods. In  Table \ref{table:max_cut_table} we give a summary of the results for 100 random non-regular unweighted graph instances with 15 to 19 vertices. We can see that our method succeeds in more instances while the overlap achieved is also much higher.

There are many reasons why non-regular random graphs are ``harder'' than regular graphs. The first is that the ground state of a regular graph, due to its symmetry, is highly degenerate where the optimiser could easily reach without ``stucking'' in a sub-optimal minimum. The second is that the Hamiltonian corresponding to a random graph has more distinct eigenvalues and as it was shown numerically by \cite{nannicini2019performance}, the number of distinct eigenvalues correlates inversely with the performance of hardware efficient ansatz.

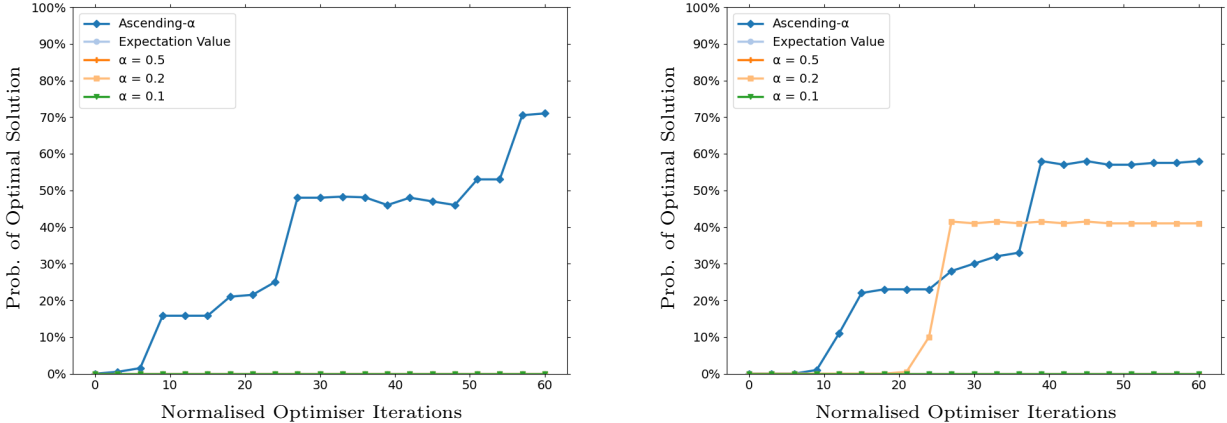
\begin{figure*}
\begin{tikzpicture}
\node (img1)  {\includegraphics[scale=0.33]{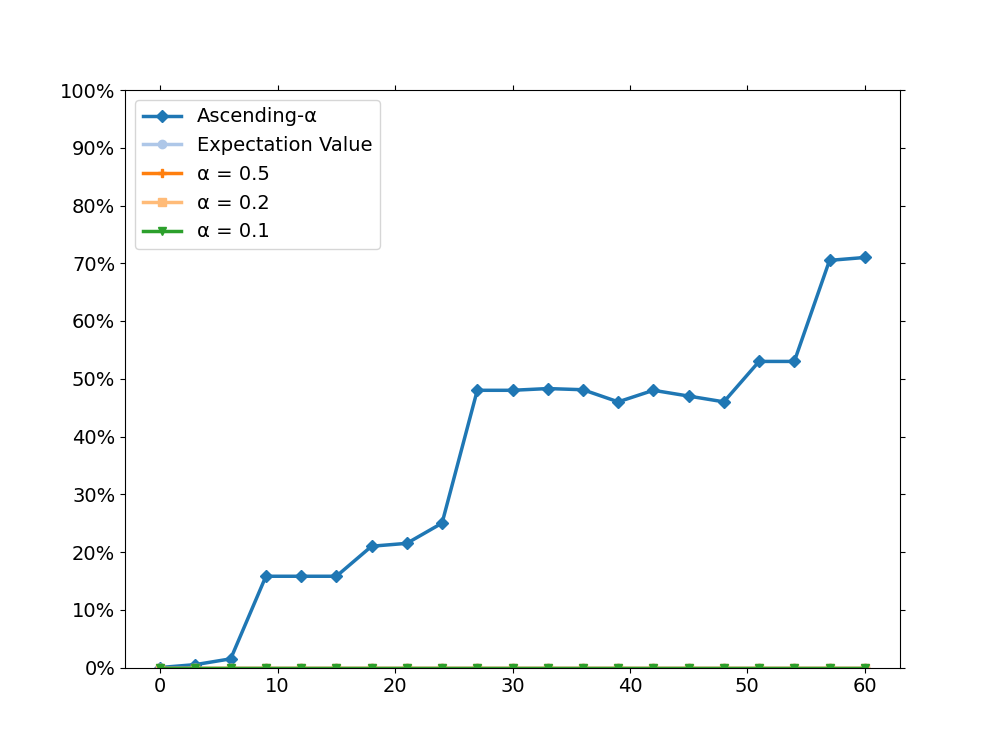}};
\node[below=of img1, node distance=0cm, yshift=1.5cm] {\scriptsize Normalised Optimiser Iterations};
\node[left=of img1, xshift=0.7cm, node distance=0cm, rotate=90, anchor=center,yshift=-0.7cm] {\scriptsize Prob. of Optimal Solution};
\node[right=of img1, xshift=-1cm] (img2)  {\includegraphics[scale=0.33]{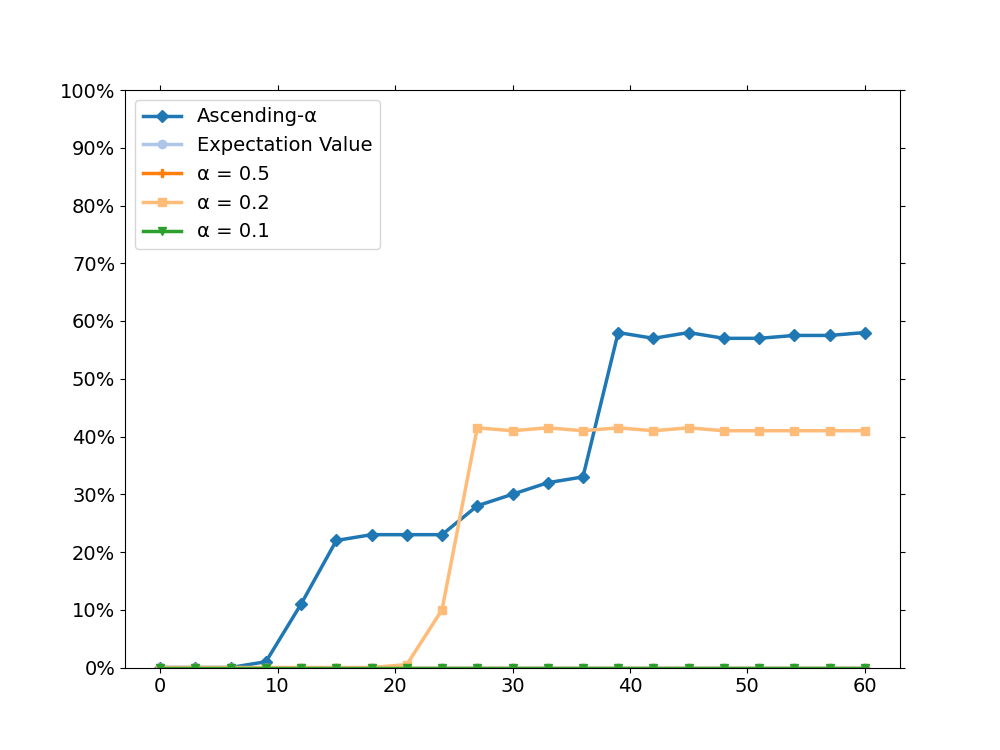}};
\node[below=of img2, node distance=0cm, yshift=1.5cm] {\scriptsize Normalised Optimiser Iterations};
\node[left=of img2, xshift=0.7cm, node distance=0cm, rotate=90, anchor=center, yshift=-0.7cm] {\scriptsize Prob. of Optimal Solution};
\end{tikzpicture}
\caption{\emph{Max-Cut} instances with 17 vertices for random non-regular unweighted graphs. \emph{Ascending-CVaR}, drawn with a blue line (diamond marker), results in a fast and high overlap with the optimal solution in contrast to constant CVaR.}
\label{fig:maxcut_optimization}
\end{figure*}

Indicatively, in Figure \ref{fig:maxcut_optimization} we plot the probability of sampling the optimal solution over the normalised number of iterations for two random graphs with 17 vertices. For the left figure, we can see how the optimiser for the Ascending-CVaR optimisation is able to to find the optimal solution in under 10 normalised iterations which by the end of the optimisation is able to increase the probability up to $70\%$. Notably the expectation value or constant CVaR completely fail. The right part of the figure, gives another example where our approach performs better. This instance constitutes an example where smaller $\alpha$'s do not lead to better performance for constant CVaR\footnote{In most cases, small $\alpha$ gives better performance, but one cannot know a-priori which is the suitable $\alpha$ in the constant CVaR case.}. We can see in the figure that while $\alpha = 0.1$ failed, $\alpha = 0.2$ was able to achieve a high quality parameterised state. This is another indication why our approach is more flexible.

\subsubsection{\texorpdfstring{CVaR$_{\alpha_t}$-QAOA}{Lg}}

Solving the Max-Cut problem using QAOA, with small depth 
circuits, does not seem a very promising approach in any of the methods considered (constant CVaR or Ascending-CVaR). In terms of speed, all methods converged equally fast but in states with small overlap with the solution (with relatively small differences within different approaches). Having said that, as explained below, our method still gives improved performance.

While CVaR$_{\alpha_t}$-VQE optimisation results in high overlap states, CVaR$_{\alpha_t}$-QAOA produces ``flat'' states, a behaviour also observed in \cite{barkoutsos2020improving}. These states have almost equal probability amplitudes to the majority of the computational basis states. For the Max-Cut problem, as noted in \cite{farhi2014quantum}, it seems that the states produced with QAOA with small $p$ result to states with energy close to the (random) initialisation point. The spread of the energies does increase with $p$, possibly leading to a state close to the ground state, but in our analysis we focused on small $p\leq 6$. Intuitively, the main reason why QAOA cannot achieve the same probability amplitudes as VQE, in the same depth, is due to having a smaller number of parameters as well as the architecture of the ansatz \cite{hadfield2019quantum}.

\begin{figure}
\begin{tikzpicture}
\node (img)  {\includegraphics[scale=0.32]{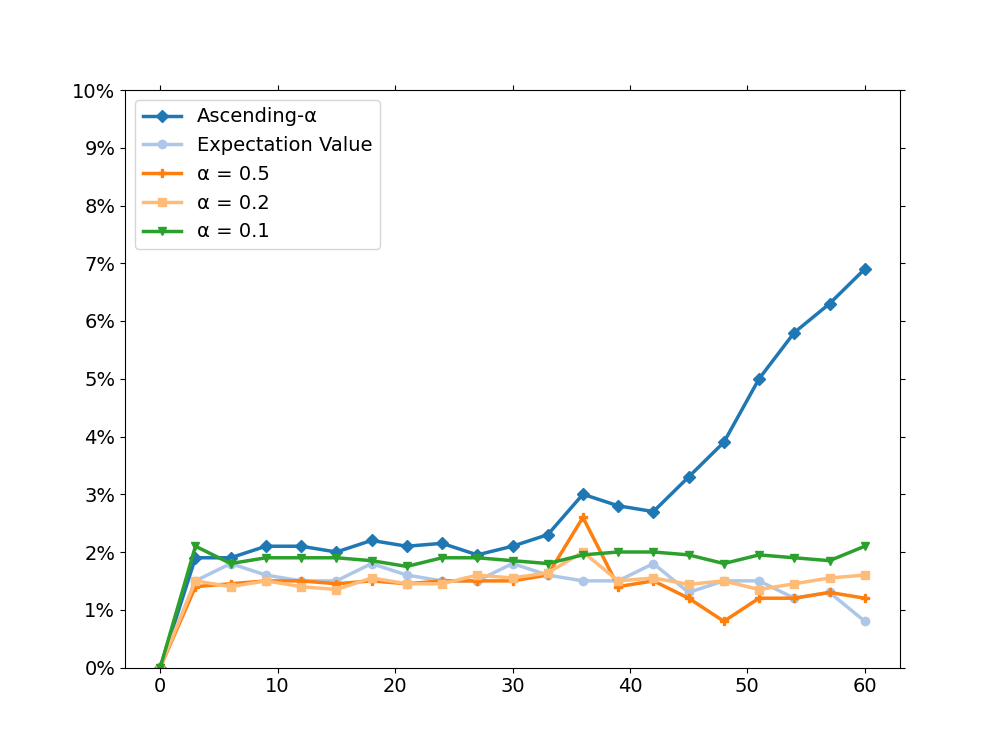}};
\node[below=of img, node distance=0cm, yshift=1.5cm] {\scriptsize Normalised Optimiser Iterations};
\node[left=of img, node distance=0cm, xshift=0.7cm, rotate=90, anchor=center,yshift=-0.7cm] {\scriptsize Prob. of Optimal Solution};
\end{tikzpicture}
\caption{CVaR$_{\alpha_t}$-QAOA optimisation with linear ascending for a Max-Cut instance of 17 qubits. The blue line (diamond marker), indicating the ascending optimisation, results in more than a 100\% increase in the overlap with the optimal solution in contrast to the expectation value or constant CVaR optimisation.}
\label{fig:qaoa_optimization}
\end{figure}

Note that the parameter space is filled with sub-optimal local minima. Constant CVaR objective functions with different confidence level $\alpha$'s lead to different energy landscape. This means that a local minimum for a confidence level $\alpha_1$ does not, in general, correspond to a local minimum for a confidence level $\alpha_2$ if $\alpha_1 \neq \alpha_2$. This is probably the reason that we get improved performance. For example, in Figure \ref{fig:qaoa_optimization} we see how Ascending-CVaR can avoid local minima. In this example all constant CVaR achieve less than $3\%$ overlap with the ground state, while the Ascending-CVaR gives $7\%$.

\subsection{Number Partitioning}

On \emph{Number Partitioning} we tested instances with 17 to 20 integers, on both VQE and QAOA. 

\subsubsection{\texorpdfstring{CVaR$_{\alpha_t}$-VQE}{Lg}}

On CVaR$_{\alpha_t}$-VQE we tested 300 instances with 17 to 20 integers, sampled randomly from three sets; $N_1 = \{0, \ldots, 200\}$, $N_2 = \{0, \ldots, 500\}$ and $N_3 = \{0, \ldots, 750\}$. We highlight that the smaller the set that the numbers are uniformly drawn from, the easier the optimiser succeeds in finding the optimal solution. A summary of the results is given at Table \ref{table:part}.

\begin{table*}[]
\begin{center}

\begin{tabular}{||c||c|c|c|c|c||c|c|c|c|c||}
    \hline
    NP&\multicolumn{5}{c||}{Successful Instances}&\multicolumn{5}{c||}{Average Overlap}\\
    \hline
     & $\alpha_t$ & 0.1 & 0.2 & 0.5 & 1 & $\alpha_t$ & 0.1 & 0.2 & 0.5 & 1 \\
     \hline
     $N_1$  & 87 & 85 & 66 & 16 & 2 & 54.17 & 11.50 & 16.56 & 7.94 & 0.99\\
     $N_2$  & 80 & 69 & 29 & 11 & 0 & 48.33 & 10.24 & 7.56 & 5.88 & 0.4 \\
     $N_3^*$  & 95 & 58 & 24 & 9 & 0 & 56.85 & 8.24 & 5.84& 3.45 & 0.16\\
     \hline
\end{tabular}
\caption{Results table for the \emph{Number Partitioning} problem (VQE) for the three different sets $N_1$, $N_2$ and $N_3^*$, where the star at the last set indicates that we used the sigmoid ascending function.}
\label{table:part}
\end{center}
\end{table*}

\begin{figure*}
\begin{tikzpicture}
\node (img1)  {\includegraphics[scale=0.33]{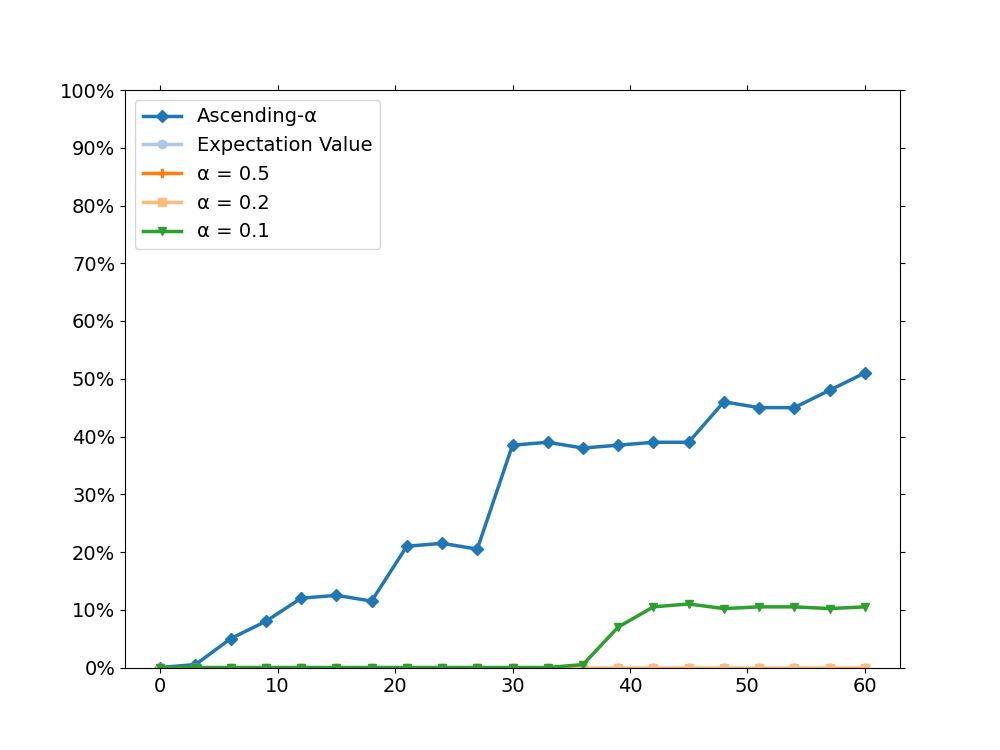}};
\node[below=of img1, node distance=0cm, yshift=1.5cm] {\scriptsize Normalised Optimiser Iterations};
\node[left=of img1, xshift=0.7cm, node distance=0cm, rotate=90, anchor=center,yshift=-0.7cm] {\scriptsize Prob. of Optimal Solution};
\node[xshift=-1cm, right=of img1] (img2)  {\includegraphics[scale=0.33]{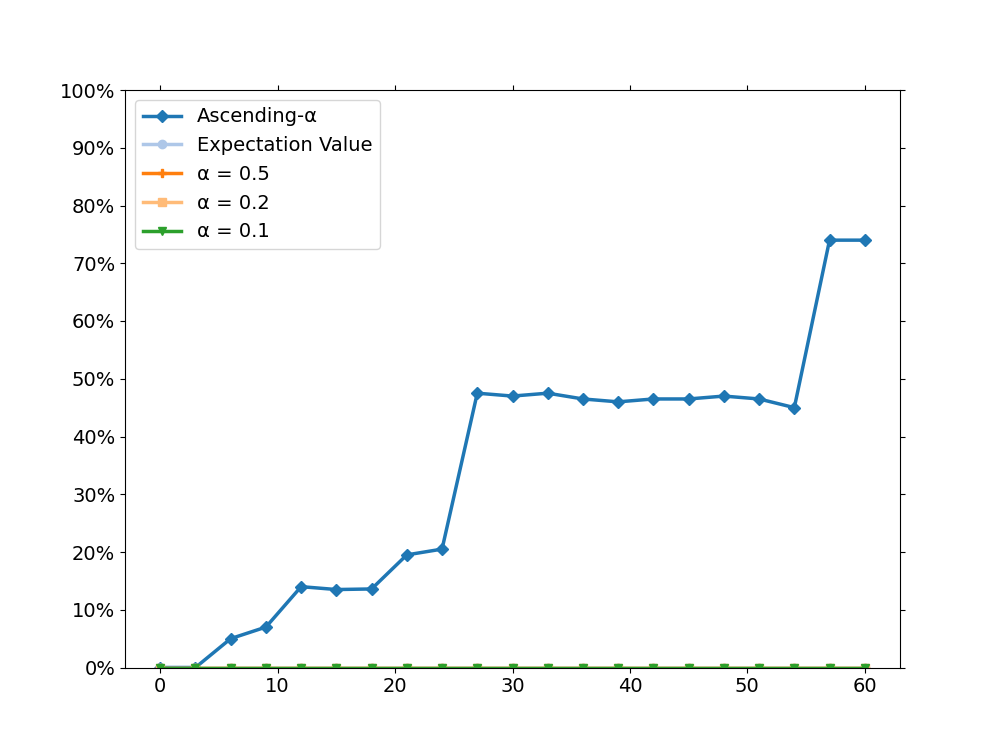}};
\node[below=of img2, node distance=0cm, yshift=1.5cm] {\scriptsize Normalised Optimiser Iterations};
\node[left=of img2, xshift=0.7cm, node distance=0cm, rotate=90, anchor=center, yshift=-0.7cm] {\scriptsize Prob. of Optimal Solution};
\end{tikzpicture}
\caption{Probability of sampling the optimal solution for \emph{Number Partitioning} instances with 17-20 integers uniformly drawn from the sets $N_1 = \{0,\ldots, 200\}$ (on the left) and $N_2 = \{0,\ldots, 500\}$ (on the right). The blue line (diamond marker), indicating \emph{Ascending-CVaR} outperforms constant CVaR in terms of speed and overlap with the optimal solution.}
\label{fig:vqe-number_part}
\end{figure*}
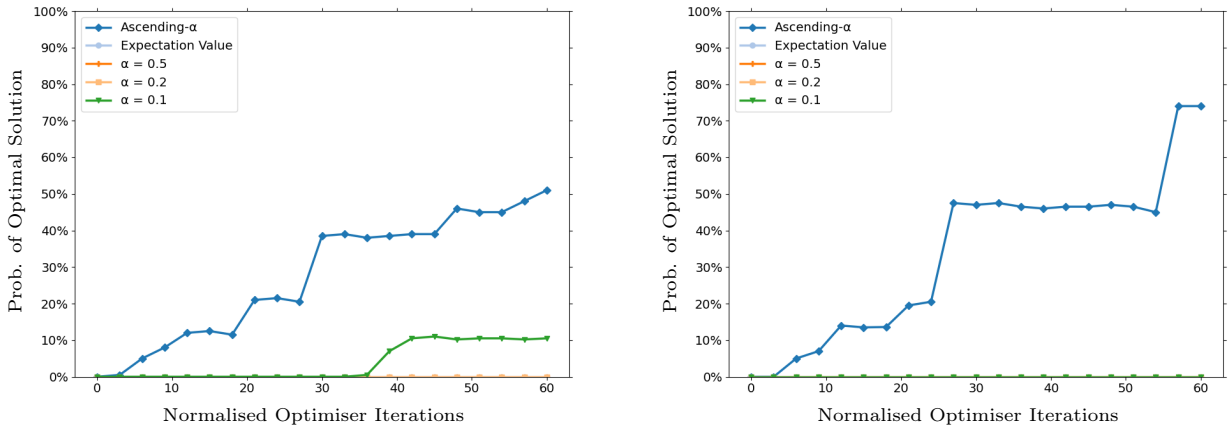

For the first two sets, we used a linear ascending function with an ascending factor $\lambda = 0.03$. Further optimisation of the parameter may lead in either faster convergence or more successful instances. Either way, the Ascending-CVaR method outperforms constant CVaR and the expectation value objective function on the aforementioned metrics (e.g. see typical performance on Figure \ref{fig:vqe-number_part}).

For the last set $N_3$, constant CVaR and the expectation value as objective functions struggled to achieve even a small overlap with the optimal solution. Indicatively, at $40\%$ of the cases none of the constant CVaR objective functions could be ``successful''\footnote{Recall, that successful in our convention, means to achieve overlap of at least $10\%$ with the optimal solution.}. We found that by choosing a sigmoid ascending function, the optimiser is able to attain a high quality parameterised state and succeed in the majority of instances ($95\%$). The trade-off is that using the sigmoid ascending function, in contrast to linear ascending, comes with some cost of more circuit shots in order to achieve the same accuracy. Note also, that the linear ascending function, while performing worse than the sigmoid, it was still more successful than the constant CVaR objective functions.

\subsubsection{\texorpdfstring{CVaR$_{\alpha_t}$-QAOA}{Lg}}

While CVaR$_{\alpha_t}$-VQE optimisation efficiently achieved a high overlap state already within the first layer for instances drawn from the two sets $N_1$ and $N_2$, CVaR$_{\alpha_t}$-QAOA failed to achieve a high overlap on small depths. To address this issue without having to increase the depth of the ansatz we chose to work on instances drawn from the smaller set $M = \{0, \ldots, 50\}$. For the Number Partitioning problem, the cost function's parameter space is highly dependent on the set we draw the numbers from. The unitary transformation $e^{i\gamma H_C}$ is composed of $e^{i\gamma n_k n_l\sigma_z^k\sigma_z^l}$ terms where $n_k, n_l$ correspond to the numbers on the $k$ and $l$ index respectively. The parameter $\gamma$ is then restricted to $0 \leq \gamma < 2\pi/(n_jn_m)$ with $n_j$ and $n_m$ corresponding to the two smallest numbers of the set.

\begin{figure}
\begin{tikzpicture}
\node (img)  {\includegraphics[scale=0.32]{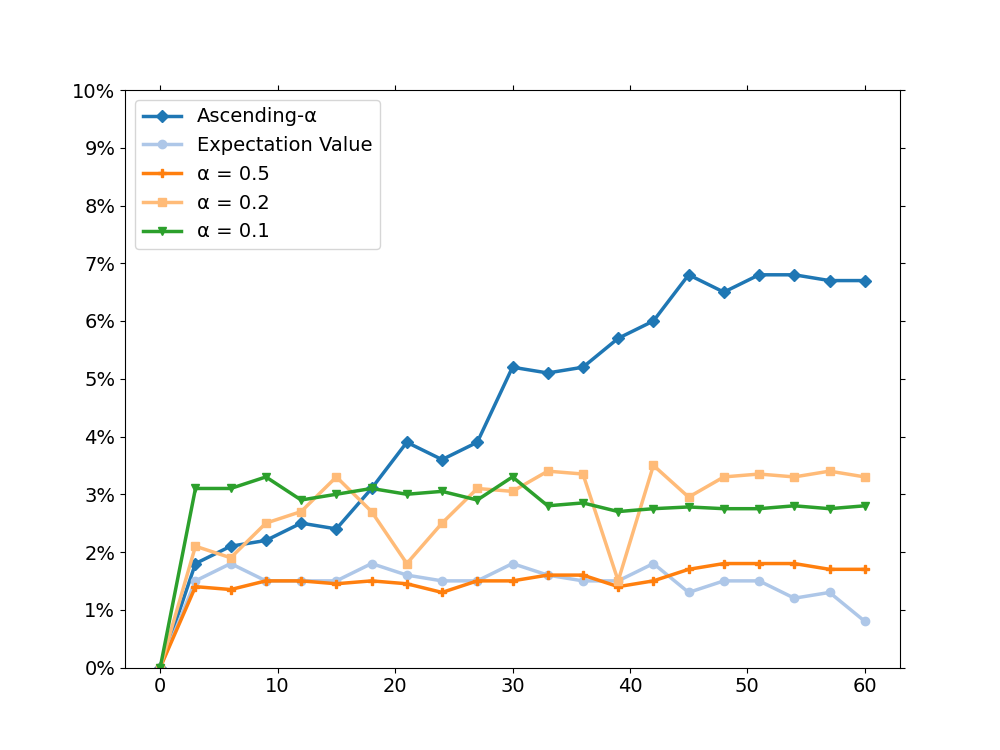}};
\node[below=of img, node distance=0cm, yshift=1.5cm] {\scriptsize Normalised Optimiser Iterations};
\node[left=of img, node distance=0cm, xshift=0.7cm, rotate=90, anchor=center,yshift=-0.7cm] {\scriptsize Prob. of Optimal Solution};
\end{tikzpicture}
\caption{CVaR$_{\alpha_t}$-QAOA for an 18-integer instance Number Partitioning problem with $p=4$. The blue line (diamond marker), indicating Ascending-CVaR optimisation, is able to achieve 100\% increase in the overlap with the optimal solution in respect to the other objective functions.}
\label{fig:qaoa-number_part}
\end{figure}

Our method succeeds in finding quantum states with higher overlap, unreachable with constant CVaR optimisation, possibly because it avoids the high amount of local minima. Indicatively in Figure \ref{fig:qaoa-number_part} we see an example where Ascending-CVaR achieves more than double overlap with the optimal solution than other methods, but is still below the threshold of $10\%$ required to classify this as a ``successful run''.

\subsection{Portfolio Optimisation}

On \emph{Portfolio Optimisation} we tested  
instances with 16 to 20 assets, on both VQE and QAOA, with a budget drawn uniformly at random from the set $B = \{0, ... n\}$ where $n$ is the number of assets and many different risk factors $q$. 

\subsubsection{\texorpdfstring{CVaR$_{\alpha_t}$-VQE}{Lg}}
\begin{figure*}
\begin{tikzpicture}
\node (img1)  {\includegraphics[scale=0.33]{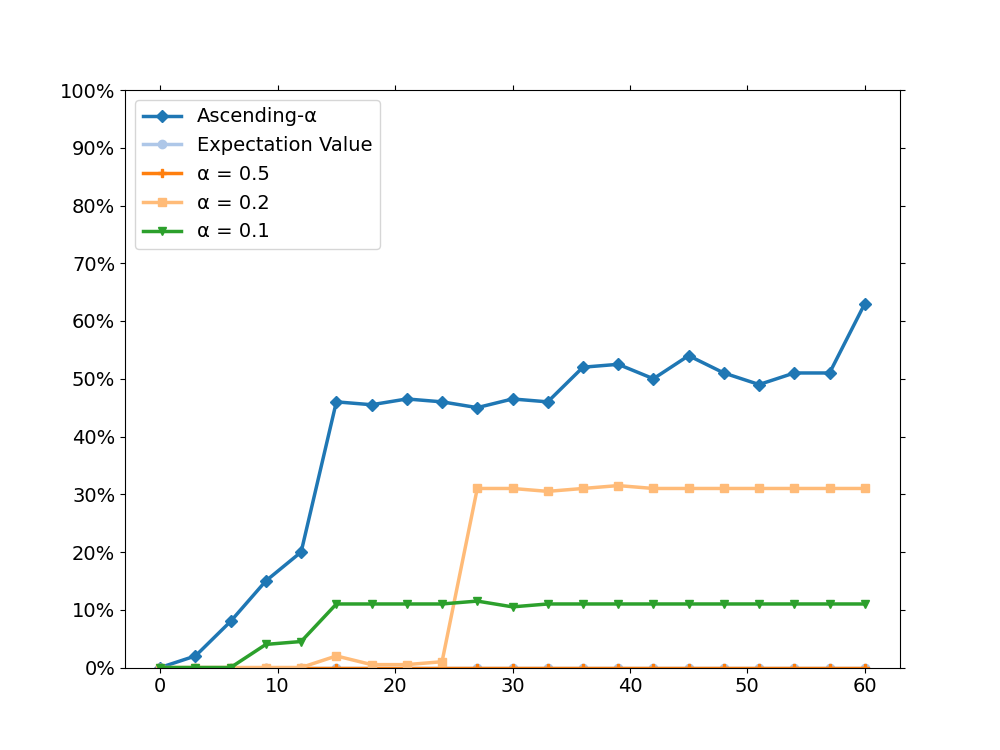}};
\node[below=of img1, node distance=0cm, yshift=1.5cm] {\scriptsize Normalised Optimiser Iterations};
\node[left=of img1, xshift=0.7cm, node distance=0cm, rotate=90, anchor=center,yshift=-0.7cm] {\scriptsize Prob. of Optimal Solution};
\node[xshift=-1cm, right=of img1] (img2)  {\includegraphics[scale=0.33]{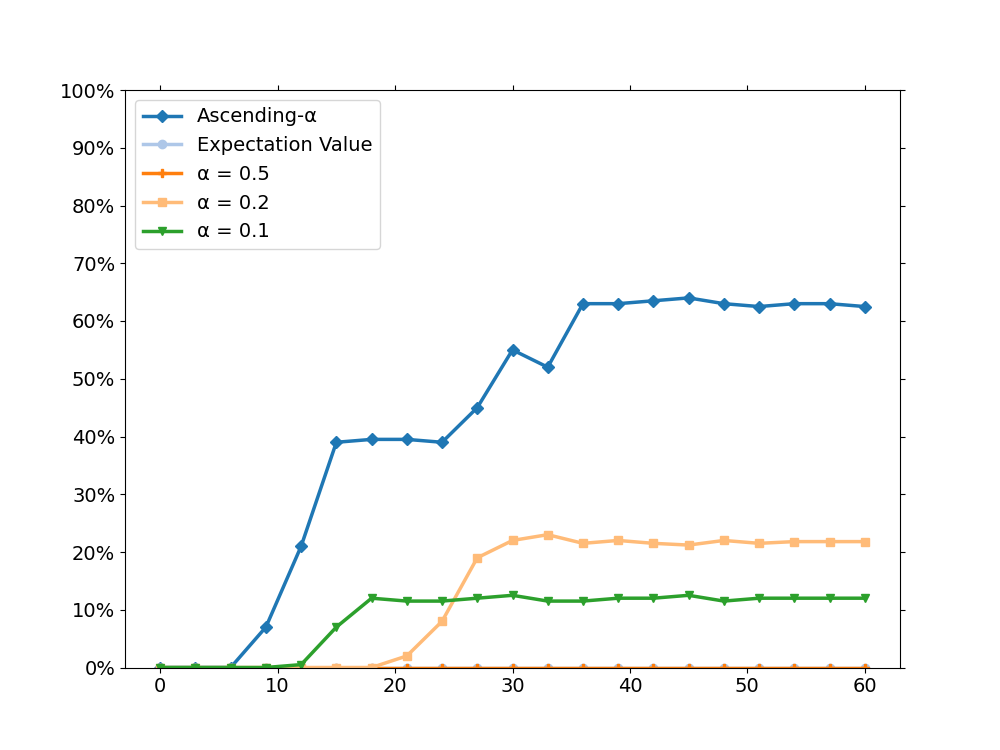}};
\node[below=of img2, node distance=0cm, yshift=1.5cm] {\scriptsize Normalised Optimiser Iterations};
\node[left=of img2, xshift=0.7cm, node distance=0cm, rotate=90, anchor=center, yshift=-0.7cm] {\scriptsize Prob. of Optimal Solution};
\end{tikzpicture}
\caption{Portfolio Optimisation problem for 18 and 20 asset instances with linear ascending and $\lambda = 0.045$. The blue line (diamond marker) indicates the CVaR$_{\alpha_t}$-VQE optimisation which already within the first 15 optimiser iterations has achieved over $40\%$ overlap, compared to constant $\alpha$'s where either fail ($\alpha = 0.5, 1$), or lead to slower and sub-optimal convergence ($\alpha = 0.1, 0.2$).}
\label{fig:porfolio_optimisation}
\end{figure*}

We used linear ascending with an ascending factor $\lambda = 0.045$ and the confidence level was initialised on $\alpha_0 = 0.01$. The results are summarised in Table \ref{table:port_opt}.
\begin{table*}
\begin{center}
\begin{tabular}{||c||c|c|c|c|c||c|c|c|c|c||}
    \hline
    Portfolio Optimisation&\multicolumn{5}{c||}{Successful Instances}&\multicolumn{5}{c||}{Average Overlap}\\
    \hline
     & $\alpha_t$ & 0.1 & 0.2 & 0.5 & 1 & $\alpha_t$ & 0.1 & 0.2 & 0.5 & 1 \\
     \hline
     Random Portfolios  & 100 & 100 & 100 & 16 & 1 & 63.25 & 13.35 & 24.74 & 9.42 & 0.64\\
     \hline
\end{tabular}
\caption{Results table for the \emph{Portfolio Optimisation} problem (VQE) for 100 random Portfolios with 16 to 20 assets.}
\label{table:port_opt}
\end{center}
\end{table*}
In Figure \ref{fig:porfolio_optimisation} we see the typical performance of two different instances  where we plotted the probability of obtaining the optimal solution over the normalised number of optimiser iterations for the CVaR$_{\alpha_t}$-VQE.

 We highlight the fact that Ascending-CVaR and constant CVaR with $\alpha = 0.1, 0.2$ succeed in achieving at least $10\%$ overlap on all instances tested (see results on Table \ref{table:port_opt}), while the expectation value ($\alpha=1)$ failed in almost all cases. Moreover, it is worth noting that our method offers a significant improvement in comparison with all the other approaches in the speed that this overlap was achieved (in terms of normalised optimiser iterations and circuit repetitions) and in the overall magnitude of the overlap achieved (see also Table \ref{table:port_opt}).

\subsubsection{\texorpdfstring{CVaR$_{\alpha_t}$-QAOA}{Lg}}

CVaR$_{\alpha_t}$-QAOA, similarly with \cite{barkoutsos2020improving}, underperforms significantly in terms of overlap with the optimal state, compared to CVaR$_{\alpha_t}$-VQE. Specifically, keeping the depth as in previous parts, and without increasing the shots each circuit is implemented, all methods fail achieving overlap with the optimal solution well below $1\%$. There are several reasons for this failure, including the \emph{Reachability Deficits} \cite{ Akshay2021reachability}, the large problem density \cite{nannicini2019performance} and \emph{Barren Plateaus} \cite{mcclean2018barren}. This, however, goes beyond the focus of this paper that is to find a way to improve the performance of previously used objective functions. To illustrate the improvement, we could have used (significantly) larger number of shots, where Ascending-CVaR would start showing better performance. This would make the comparison with other problems unfair (where in all cases we used the same ``normalised'' number of shots), and it would still not present a practical way to solve the Portfolio Optimisation problem (VQE is much better), so we omitted it.

\section{Conclusions}
\label{conc}

We introduced a novel type of objective function, Ascending-CVaR, to be used in variational quantum algorithms for any combinatorial optimisation problem. The starting point is the (constant) CVaR objective function of \cite{barkoutsos2020improving}, where they illustrated that for any choice of risk $\alpha$ the true ground state is a minimum, and that with (typically small values of) $\alpha$ one can improve the performance compared to the ``standard'' expectation-value objective function. Our idea was to use an evolving objective function that ``passes'' through all the different values of $\alpha$ to finish at the expectation value. This, intuitively, avoids getting stuck at local minima since the energy landscape for different $\alpha$'s differs apart from the global minimum.

We tested numerically the proposal on three combinatorial optimisation problems (Max-Cut, Number Partitioning and Portfolio Optimisation), where in agreement with prior works we found that for these problems VQE seems more promising than QAOA with small depth. The improvement that Ascending-CVaR provides to VQE and QAOA are similar but we focus on VQE here since this was the overall more promising approach to solve the corresponding optimisation problems.

We observed that Ascending-CVaR gave much greater on average overlap with the optimal solution (see Table \ref{table-overview}). In Portfolio Optimisation and Number Partitioning we got 10 times greater overlap than the expectation value (while we got at least double overlap than the best constant CVaR choice). In Max-Cut we got smaller improvement ($80\%$) compared to the expectation value, but note that the constant CVaR actually gave much smaller overlap. Perhaps the most important feature is that in the Number Partitioning and Max-Cut, Ascending-CVaR succeeded in finding the solution in many instances that no other approach achieved more than the small chosen threshold of $10\%$ overlap. This indicates that not only the approach improves the quality of the results, but is plausible that instances that are believed to be ``hard'' with the other methods, will become ``easy'' and thus solvable. 

\begin{table*}
\begin{center}
\begin{tabular}{||c||c|c|c|c||c|c|c||}
    \hline
    Problem Type &\multicolumn{4}{c||}{Successful Instances $\%$} &\multicolumn{3}{c||}{Overlap Improvement $\%$}\\
    \hline
    & $\alpha_t$ & 0.1 & 0.2 & Expectation Value & $\alpha_t$ & 0.1 & 0.2\\
     \hline
 Portfolio Optimisation & 100 & 100& 100&  1 & 9782 & 2125 & 4023\\
     \hline
 Number Partitioning & 87.3 & 70 & 39.6 & 0.6 & 10225 & 1858 & 1464\\
    \hline
 Max-Cut & 96 & 84 & 81 & 53 & 78.5 & -66.52 & -40.8 \\
     \hline
\end{tabular}
\caption{Overview of our method}
\label{table-overview}
\end{center}
\end{table*}

Beyond the accuracy of the result, another factor to evaluate the performance of variational quantum algorithms is the speed, that can be counted with respect to the (average) number of iterations the optimisation needs to run until the algorithm outputs a (candidate) solution. Since our proposal ``passes'' through several choices of $\alpha$, one could expect that the ``trade-off'' for better overlap would be slower speed and thus more optimisation iterations. Interestingly, not only we do not get any cost in speed, but in most cases we see an improvement, i.e. our method requires fewer iterations to reach the threshold of $10\%$ overlap with the solution (see Table \ref{table:speed}). The only case that our method required slightly more iterations than the $\alpha=0.1$ was for the case that we actually observed the greater improvement in overlap. This was the Number Partitioning from the set $N_3$, where the overlap was seven times better than the ``next best'' case, and 350 times greater overlap than the expectation value (see Table \ref{table:part}).

\begin{table*}[]
\begin{center}

\begin{tabular}{||c||c||c|c|c||}
    \hline
    Problem Type & Instance Class &\multicolumn{3}{c||}{Average Normalised Iterations}\\
    \hline
    & & $\alpha_t$ & 0.1 & 0.2\\
     \hline
 Portfolio Optimisation &    Random Portfolios &9.64 & 11.13 & 16.29 \\
     \hline
 Number Partitioning &  $N_1$ & 12.1 & 19.76 & 25.09 \\
    \hline
    &$N_2$ & 14.73 & 24.13 & 28.86 \\
     \hline
 & $N_3^*$ & 27.12 & 25.12 & 33.62 \\
     \hline
 Max-Cut &   Random Graphs & 8.75 & 9.3 & 10.8 \\
     \hline
\end{tabular}
\caption{Average Normalised Optimiser Iterations to achieve at least $10\%$ overlap with the optimal solution for the three different combinatorial problems.}
\label{table:speed}
\end{center}
\end{table*}

Our work, not only offers a generic method to improve the performance of variational quantum algorithms for combinatorial optimisation problems, it also suggests a new direction of research where dynamic objective functions can be used to boost the performance in terms of quality and speed of near-term quantum algorithms. An immediate follow up to the proposal suggested here is to generalise our approach.
Concretely, our method introduces two extra degrees of freedom. The hyperparameter $\lambda$ and the function according to which the parameter $\alpha$ increases. It is worth exploring a more systematic rule on how to fix these degrees of freedom according to the problem considered and the features of the specific instance. Finally, considering other dynamic objective functions is another direction that is worth pursuing. 

\textbf{The code for the experiments is available at \href{https://github.com/ioankolot/ascending_cvar}{GitHub} \footnote{\url{https://github.com/ioankolot/ascending_cvar}}.}

\section*{Acknowledgements}

P.W. acknowledges support by the UK Hub in Quantum Computing and Simulation, part of the UK National Quantum Technologies Programme with funding from UKRI EPSRC grant EP/T001062/1 and the  Collaborative Computational Project - Quantum Computing (CCP-QC) with funding from UKRI EPSRC grant EP/T026715/1.

\newpage
\bibliographystyle{unsrt}
\bibliography{References}

\begin{thebibliography}{10}

\bibitem{preskill2018quantum}
John Preskill.
\newblock Quantum computing in the nisq era and beyond.
\newblock {\em Quantum}, 2:79, 2018.

\bibitem{arute2019quantum}
Frank Arute, Kunal Arya, Ryan Babbush, Dave Bacon, Joseph~C Bardin, Rami
  Barends, Rupak Biswas, Sergio Boixo, Fernando~GSL Brandao, David~A Buell,
  et~al.
\newblock Quantum supremacy using a programmable superconducting processor.
\newblock {\em Nature}, 574(7779):505--510, 2019.

\bibitem{egger2020warm}
Daniel~J Egger, Jakub Marecek, and Stefan Woerner.
\newblock Warm-starting quantum optimization.
\newblock {\em arXiv preprint arXiv:2009.10095}, 2020.

\bibitem{bravyi2019obstacles}
Sergey Bravyi, Alexander Kliesch, Robert Koenig, and Eugene Tang.
\newblock Obstacles to state preparation and variational optimization from
  symmetry protection.
\newblock {\em arXiv preprint arXiv:1910.08980}, 2019.

\bibitem{skolik2021layerwise}
Andrea Skolik, Jarrod~R McClean, Masoud Mohseni, Patrick van~der Smagt, and
  Martin Leib.
\newblock Layerwise learning for quantum neural networks.
\newblock {\em Quantum Machine Intelligence}, 3(1):1--11, 2021.

\bibitem{nakanishi2020sequential}
Ken~M Nakanishi, Keisuke Fujii, and Synge Todo.
\newblock Sequential minimal optimization for quantum-classical hybrid
  algorithms.
\newblock {\em Physical Review Research}, 2(4):043158, 2020.

\bibitem{wauters2020reinforcement}
Matteo~M Wauters, Emanuele Panizon, Glen~B Mbeng, and Giuseppe~E Santoro.
\newblock Reinforcement learning assisted quantum optimization.
\newblock {\em arXiv preprint arXiv:2004.12323}, 2020.

\bibitem{li2020quantum}
Li~Li, Minjie Fan, Marc Coram, Patrick Riley, Stefan Leichenauer, et~al.
\newblock Quantum optimization with a novel gibbs objective function and ansatz
  architecture search.
\newblock {\em Physical Review Research}, 2(2):023074, 2020.

\bibitem{benedetti2019generative}
Marcello Benedetti, Delfina Garcia-Pintos, Oscar Perdomo, Vicente
  Leyton-Ortega, Yunseong Nam, and Alejandro Perdomo-Ortiz.
\newblock A generative modeling approach for benchmarking and training shallow
  quantum circuits.
\newblock {\em npj Quantum Information}, 5(1):1--9, 2019.

\bibitem{cheng2018information}
Song Cheng, Jing Chen, and Lei Wang.
\newblock Information perspective to probabilistic modeling: Boltzmann machines
  versus born machines.
\newblock {\em Entropy}, 20(8):583, 2018.

\bibitem{barkoutsos2020improving}
Panagiotis~Kl Barkoutsos, Giacomo Nannicini, Anton Robert, Ivano Tavernelli,
  and Stefan Woerner.
\newblock Improving variational quantum optimization using cvar.
\newblock {\em Quantum}, 4:256, 2020.

\bibitem{albash2018adiabatic}
Tameem Albash and Daniel~A Lidar.
\newblock Adiabatic quantum computation.
\newblock {\em Reviews of Modern Physics}, 90(1):015002, 2018.

\bibitem{cerezo2020variational}
Marco Cerezo, Andrew Arrasmith, Ryan Babbush, Simon~C Benjamin, Suguru Endo,
  Keisuke Fujii, Jarrod~R McClean, Kosuke Mitarai, Xiao Yuan, Lukasz Cincio,
  et~al.
\newblock Variational quantum algorithms.
\newblock {\em arXiv preprint arXiv:2012.09265}, 2020.

\bibitem{Sack2021quantumannealing}
Stefan~H. Sack and Maksym Serbyn.
\newblock Quantum annealing initialization of the quantum approximate
  optimization algorithm.
\newblock {\em {Quantum}}, 5:491, July 2021.

\bibitem{brandao2018fixed}
Fernando~GSL Brandao, Michael Broughton, Edward Farhi, Sam Gutmann, and Hartmut
  Neven.
\newblock For fixed control parameters the quantum approximate optimization
  algorithm's objective function value concentrates for typical instances.
\newblock {\em arXiv preprint arXiv:1812.04170}, 2018.

\bibitem{lucas2014ising}
Andrew Lucas.
\newblock Ising formulations of many np problems.
\newblock {\em Frontiers in Physics}, 2:5, 2014.

\bibitem{farhi2014quantum}
Edward Farhi, Jeffrey Goldstone, and Sam Gutmann.
\newblock A quantum approximate optimization algorithm.
\newblock {\em arXiv preprint arXiv:1411.4028}, 2014.

\bibitem{wang2018quantum}
Zhihui Wang, Stuart Hadfield, Zhang Jiang, and Eleanor~G Rieffel.
\newblock Quantum approximate optimization algorithm for maxcut: A fermionic
  view.
\newblock {\em Physical Review A}, 97(2):022304, 2018.

\bibitem{crooks2018performance}
Gavin~E Crooks.
\newblock Performance of the quantum approximate optimization algorithm on the
  maximum cut problem.
\newblock {\em arXiv preprint arXiv:1811.08419}, 2018.

\bibitem{moussa2020quantum}
Charles Moussa, Henri Calandra, and Vedran Dunjko.
\newblock To quantum or not to quantum: towards algorithm selection in
  near-term quantum optimization.
\newblock {\em Quantum Science and Technology}, 5(4):044009, 2020.

\bibitem{farhi2019quantum}
Edward Farhi, Jeffrey Goldstone, Sam Gutmann, and Leo Zhou.
\newblock The quantum approximate optimization algorithm and the
  sherrington-kirkpatrick model at infinite size.
\newblock {\em arXiv preprint arXiv:1910.08187}, 2019.

\bibitem{mertens2006number}
Stephan Mertens.
\newblock Number partitioning.
\newblock {\em Computational Complexity and Statistical Physics}, page 125,
  2006.

\bibitem{korf2009multi}
Richard~E Korf.
\newblock Multi-way number partitioning.
\newblock In {\em IJCAI}, volume~9, pages 538--543, 2009.

\bibitem{orus2019quantum}
Roman Orus, Samuel Mugel, and Enrique Lizaso.
\newblock Quantum computing for finance: Overview and prospects.
\newblock {\em Reviews in Physics}, 4:100028, 2019.

\bibitem{venturelli2019reverse}
Davide Venturelli and Alexei Kondratyev.
\newblock Reverse quantum annealing approach to portfolio optimization
  problems.
\newblock {\em Quantum Machine Intelligence}, 1(1):17--30, 2019.

\bibitem{kellerer2000selecting}
Hans Kellerer, Renata Mansini, and M~Grazia Speranza.
\newblock Selecting portfolios with fixed costs and minimum transaction lots.
\newblock {\em Annals of Operations Research}, 99(1):287--304, 2000.

\bibitem{egger2020quantum}
Daniel~J Egger, Claudio Gambella, Jakub Marecek, Scott McFaddin, Martin
  Mevissen, Rudy Raymond, Andrea Simonetto, Stefan Woerner, and Elena Yndurain.
\newblock Quantum computing for finance: state of the art and future prospects.
\newblock {\em IEEE Transactions on Quantum Engineering}, 2020.

\bibitem{cohen2020portfolio}
Jeffrey Cohen, Alex Khan, and Clark Alexander.
\newblock Portfolio optimization of 40 stocks using the dwave quantum annealer.
\newblock {\em arXiv preprint arXiv:2007.01430}, 2020.

\bibitem{Slate2021quantumwalkbased}
N.~Slate, E.~Matwiejew, S.~Marsh, and J.~B. Wang.
\newblock Quantum walk-based portfolio optimisation.
\newblock {\em {Quantum}}, 5:513, July 2021.

\bibitem{mugel2020dynamic}
Samuel Mugel, Carlos Kuchkovsky, Escolastico Sanchez, Samuel Fernandez-Lorenzo,
  Jorge Luis-Hita, Enrique Lizaso, and Roman Orus.
\newblock Dynamic portfolio optimization with real datasets using quantum
  processors and quantum-inspired tensor networks.
\newblock {\em arXiv preprint arXiv:2007.00017}, 2020.

\bibitem{more2009benchmarking}
Jorge~J Mor{\'e} and Stefan~M Wild.
\newblock Benchmarking derivative-free optimization algorithms.
\newblock {\em SIAM Journal on Optimization}, 20(1):172--191, 2009.

\bibitem{powell1994direct}
Michael~JD Powell.
\newblock A direct search optimization method that models the objective and
  constraint functions by linear interpolation.
\newblock In {\em Advances in optimization and numerical analysis}, pages
  51--67. Springer, 1994.

\bibitem{nannicini2019performance}
Giacomo Nannicini.
\newblock Performance of hybrid quantum-classical variational heuristics for
  combinatorial optimization.
\newblock {\em Physical Review E}, 99(1):013304, 2019.

\bibitem{hagberg2008exploring}
Aric Hagberg, Pieter Swart, and Daniel S~Chult.
\newblock Exploring network structure, dynamics, and function using networkx.
\newblock Technical report, Los Alamos National Lab.(LANL), Los Alamos, NM
  (United States), 2008.

\bibitem{hadfield2019quantum}
Stuart Hadfield, Zhihui Wang, Bryan O’Gorman, Eleanor~G Rieffel, Davide
  Venturelli, and Rupak Biswas.
\newblock From the quantum approximate optimization algorithm to a quantum
  alternating operator ansatz.
\newblock {\em Algorithms}, 12(2):34, 2019.

\bibitem{Akshay2021reachability}
V.~Akshay, H.~Philathong, I.~Zacharov, and J.~Biamonte.
\newblock Reachability {D}eficits in {Q}uantum {A}pproximate {O}ptimization of
  {G}raph {P}roblems.
\newblock {\em {Quantum}}, 5:532, August 2021.

\bibitem{mcclean2018barren}
Jarrod~R McClean, Sergio Boixo, Vadim~N Smelyanskiy, Ryan Babbush, and Hartmut
  Neven.
\newblock Barren plateaus in quantum neural network training landscapes.
\newblock {\em Nature communications}, 9(1):1--6, 2018.

\bibitem{peruzzo2014variational}
Alberto Peruzzo, Jarrod McClean, Peter Shadbolt, Man-Hong Yung, Xiao-Qi Zhou,
  Peter~J Love, Al{\'a}n Aspuru-Guzik, and Jeremy~L O’brien.
\newblock A variational eigenvalue solver on a photonic quantum processor.
\newblock {\em Nature communications}, 5(1):1--7, 2014.

\bibitem{moll2018quantum}
Nikolaj Moll, Panagiotis Barkoutsos, Lev~S Bishop, Jerry~M Chow, Andrew Cross,
  Daniel~J Egger, Stefan Filipp, Andreas Fuhrer, Jay~M Gambetta, Marc Ganzhorn,
  et~al.
\newblock Quantum optimization using variational algorithms on near-term
  quantum devices.
\newblock {\em Quantum Science and Technology}, 3(3):030503, 2018.

\bibitem{kandala2017hardware}
Abhinav Kandala, Antonio Mezzacapo, Kristan Temme, Maika Takita, Markus Brink,
  Jerry~M Chow, and Jay~M Gambetta.
\newblock Hardware-efficient variational quantum eigensolver for small
  molecules and quantum magnets.
\newblock {\em Nature}, 549(7671):242--246, 2017.

\bibitem{lee2018generalized}
Joonho Lee, William~J Huggins, Martin Head-Gordon, and K~Birgitta Whaley.
\newblock Generalized unitary coupled cluster wave functions for quantum
  computation.
\newblock {\em Journal of chemical theory and computation}, 15(1):311--324,
  2018.

\bibitem{wecker2018towards}
Dave Wecker, Matthew Hastings, and Matthias Troyer.
\newblock Towards practical quantum variational algorithms.
\newblock 2018.

\bibitem{zhou2020quantum}
Leo Zhou, Sheng-Tao Wang, Soonwon Choi, Hannes Pichler, and Mikhail~D Lukin.
\newblock Quantum approximate optimization algorithm: Performance, mechanism,
  and implementation on near-term devices.
\newblock {\em Physical Review X}, 10(2):021067, 2020.

\bibitem{bittel2021training}
Lennart Bittel and Martin Kliesch.
\newblock Training variational quantum algorithms is np-hard--even for
  logarithmically many qubits and free fermionic systems.
\newblock {\em arXiv preprint arXiv:2101.07267}, 2021.

\bibitem{shaydulin2019multistart}
Ruslan Shaydulin, Ilya Safro, and Jeffrey Larson.
\newblock Multistart methods for quantum approximate optimization.
\newblock In {\em 2019 IEEE High Performance Extreme Computing Conference
  (HPEC)}, pages 1--8. IEEE, 2019.

\end{thebibliography}

\appendix

\section{Variational Quantum Algorithms}

In this section we introduce the details of the two main variational quantum algorithms used in this paper.

\subsubsection{Variational Quantum Eigensolver}

The Variational Quantum Eigensolver, as proposed by \cite{peruzzo2014variational}, is a hybrid quantum-classical algorithm, originally designed to solve quantum chemistry problems, but it can be used to tackle optimisation problems \cite{nannicini2019performance}. The main idea is to a map the optimisation problem into a cost function that is translated into an interacting qubit Hamiltonian \cite{moll2018quantum}, whose ground state corresponds to the solution of the optimisation problem. 

The encoded qubit Hamiltonian, $H_C$, is decomposed into a linear combination of Pauli strings $P_a$, consisted of tensor products of Pauli Matrices $\hat{\sigma}^x, \hat{\sigma}^y, \hat{\sigma}^z$ :

\begin{equation}
    H_C = \sum_{k=1}^Mc_kP_k
    \label{vqe:decomp}
\end{equation}

where $M=\mathcal{O}(poly(n))$, $n$ is the system size and $c_k$ is the complex coefficient of the $P_k$ Pauli string. However, for combinatorial optimisation problems where the Hamiltonian is a diagonal matrix, $H_C$ is decomposed only on Pauli strings consisting of $\sigma_i^z$ operators.

VQE is initialised by creating a parameterised state $\ket{\psi(\boldsymbol{\theta})}$ whose parameters are iteratively updated by a classical optimiser so as to minimize an objective function, usually the expectation value of (\ref{vqe:decomp}). The parameterised state is created by choosing a variational form $U(\boldsymbol{\theta})$  which acts on the initial state $\ket{0}^{\otimes n}$ and produces $\ket{\psi(\boldsymbol{\theta})}$.

Our choice of variational form $U(\boldsymbol{\theta})$ is a \emph{hardware efficient ansatz} \cite{barkoutsos2020improving, kandala2017hardware}, where the qubits are initialised in the $\ket{0}$ state and $R_y(\theta_i)$-rotations are applied in each qubit along with control-Z operators. Each layer of the variational form consists of $(CZ)_{ij}$ operations with $i$ the control qubit and $j$ the target qubit, as long as the condition $i<j$ holds, and $R_y(\theta_i)$ rotations for every qubit. If $p$ is the number of layers, then the number of parameters is linear, $\mu = n\left(1+p\right)$, in the number of qubits and the variational form spans every basis state already within the first layer.

\begin{center}
    \begin{figure}
    \includegraphics[width=0.55\textwidth]{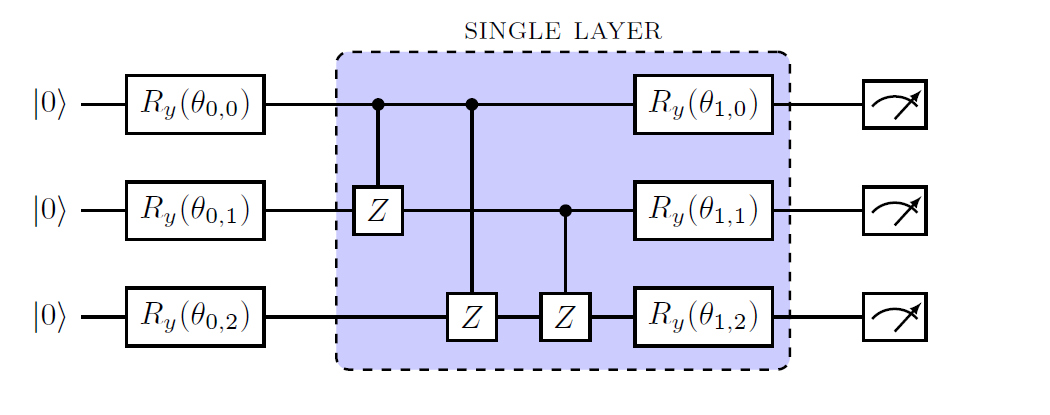}
    \caption{Single Layer Hardware Efficient Ansatz for $3$ qubits.}
    \end{figure}
\end{center}

\vspace{-8mm}
The \emph{hardware efficient ansatz} falls in the more general category of \emph{problem-agnostic ansatze}, meaning that the structure of the ansatz carries no information about the problem itself and is mostly suited for optimisation problems. Other problems use different ansatz families, like the \emph{Unitary Coupled Cluster} which is widely used in chemistry to obtain the ground state of a molecule \cite{lee2018generalized} or the \emph{Variational Hamiltonian Ansatz} which encodes the problem's Hamiltonian \cite{wecker2018towards}.
\vspace{-4mm}
\subsubsection{Quantum Approximate Optimisation Algorithm}

The \emph{Quantum Approximate Optimisation Algorithm} (QAOA) \cite{farhi2014quantum} is a variational quantum algorithm mostly used in combinatorial optimisation problems, and while in shallow depths it is analytically and numerically explored for some problems \cite{wang2018quantum, zhou2020quantum}, its performance in intermediate depths is still unknown. 

\begin{center}
    \begin{figure}
    \includegraphics[width=0.5\textwidth]{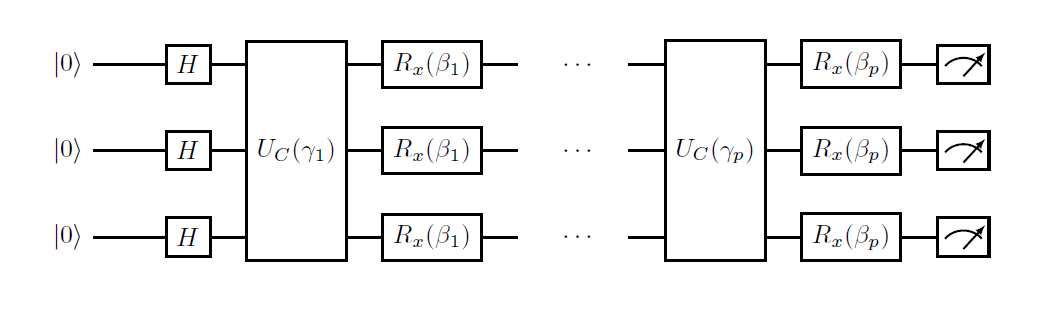}
    \caption{General framework of a $p$-layer QAOA consisting of $2p$ variational angles.}
    \label{fig:qaoacirc}
    \end{figure}
\end{center}

The QAOA algorithm applies an alternation of two unitary transformations, one encoding the cost function $H_C$, $U(H_C)=e^{-i\gamma H_C}$, and the other a mixer Hamiltonian $H_B = \sum\sigma_i^x$, $U(H_B)=e^{-i\beta H_B}$, where $\gamma$ and $\beta$ are \emph{variational angles} specifying the ``time'' for which the unitary transformations are applied. The system is initialised at the ground state of $H_B$ and the alternating ansatz of $U(H_B)$ $U(H_C)$ is applied $p$-times, with $p$ defining the number of layers of the algorithm (see Figure \ref{fig:qaoacirc}), producing the state:

\begin{equation}
	\ket{\boldsymbol{\beta}, \boldsymbol{\gamma}} = e^{-i\beta_p H_B}e^{-i\gamma_p H_C}\ldots e^{-i\beta_1 H_B}e^{-i\gamma_1 H_C}\ket{+}
\end{equation}
where $\ket{+}$ is the uniform superposition state, $\boldsymbol{\gamma} = \left(\gamma_1\ldots,\gamma_p\right)$ and $\boldsymbol{\beta} = \left(\beta_1\ldots,\beta_p\right)$. 

With sufficient repetitions of the algorithm, the expectation value is calculated as:
\begin{equation}
	F_p(\boldsymbol{\beta}, \boldsymbol{\gamma}) = \bra{\boldsymbol{\beta},\boldsymbol{\gamma}}H_C\ket{\boldsymbol{\beta}, \boldsymbol{\gamma}}
\end{equation}
until the $2p$ optimal parameters $(\boldsymbol{\beta}^*,\boldsymbol{\gamma}^*)$ are found.

If $C_{opt}$ is the optimal cost function, then the target of the algorithm is to maximise the approximation ratio, defined as:
\begin{equation}
	r^* = \frac{F_p(\boldsymbol{\beta} ^*, \boldsymbol{\gamma}^*)}{C_{opt}}
	\label{eq:approx_ratio}
\end{equation}

Finding the optimal parameters is far from trivial since the expectation value landscape is highly non-convex, filled with local minima where a classical optimiser could easily get stuck.

The hardest part of QAOA, and in general of a variational quantum algorithm, is finding the optimal parameters that will lead in a high overlap with the optimal (or near-optimal) bit-string or low expectation value. Recently, \cite{bittel2021training} proved that training the optimization parameters is NP-Hard and that the landscape of the objective function is filled with far-from-optimal local minima. One way to avoid ``getting stuck'' in a local minima is using multi-start methods \cite{shaydulin2019multistart} or heuristic methods like using the global optimum of one layer, in QAOA, as a starting point for the next \cite{zhou2020quantum}.

\section{Circuit Repetitions}
\label{appendix}

In this section, we demonstrate how our method outperforms, in terms of real circuit repetitions and quality of the output state, the previously used objective functions. We set our ``default'' circuit repetitions to $K=1000$ which we then scale it up, along the discretely increasing $\alpha$ using the expression $K/\alpha_t$, for each given time. While one may think that this would weaken our results, as illustrated below, it seems that in terms of circuit repetitions our method converges to the chosen threshold of $10\%$ faster than the best of constant CVaR or the expectation value approaches.

\begin{figure}
\begin{tikzpicture}
\node (img)  {\includegraphics[scale=0.32]{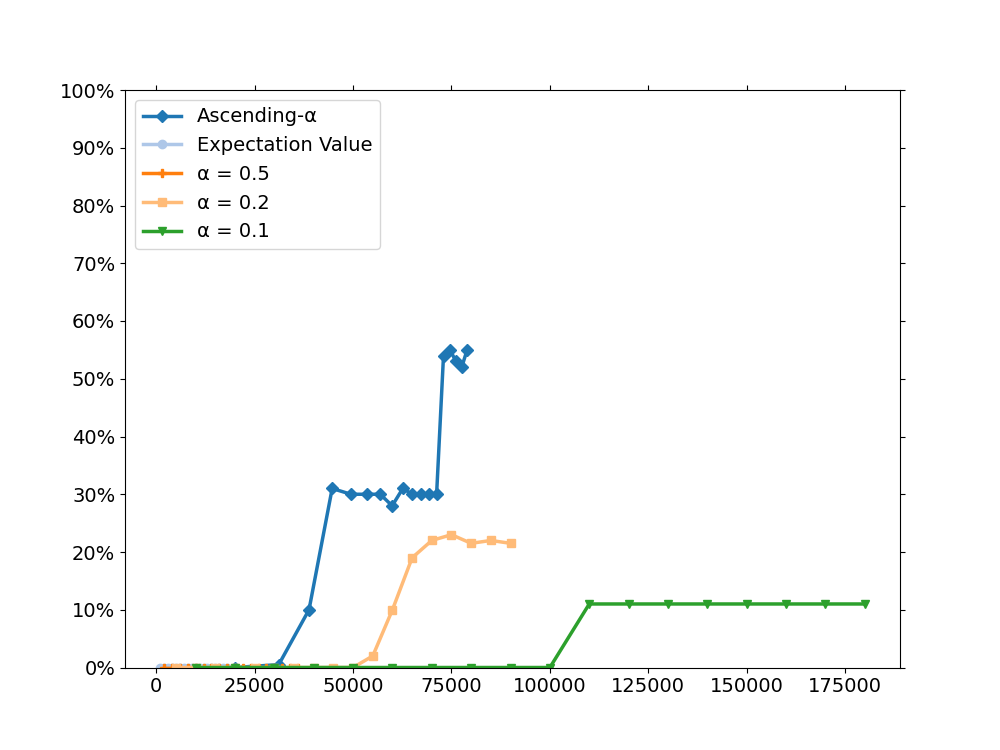}};
\node[below=of img, node distance=0cm, yshift=1.5cm] {\scriptsize Circuit Repetitions};
\node[left=of img, node distance=0cm, xshift=0.7cm, rotate=90, anchor=center,yshift=-0.7cm] {\scriptsize Prob. of Optimal Solution};
\end{tikzpicture}
    \caption{Probability of sampling an optimal solution over the circuit repetitions for a Number-Partitioning instance.}
    \label{circuit:repet}
\end{figure}

\section{Numerical analysis of Ascending factor}
\label{appendix:numerical_analysis}

\begin{figure*}
\begin{tikzpicture}
\node (img1)  {\includegraphics[scale=0.32]{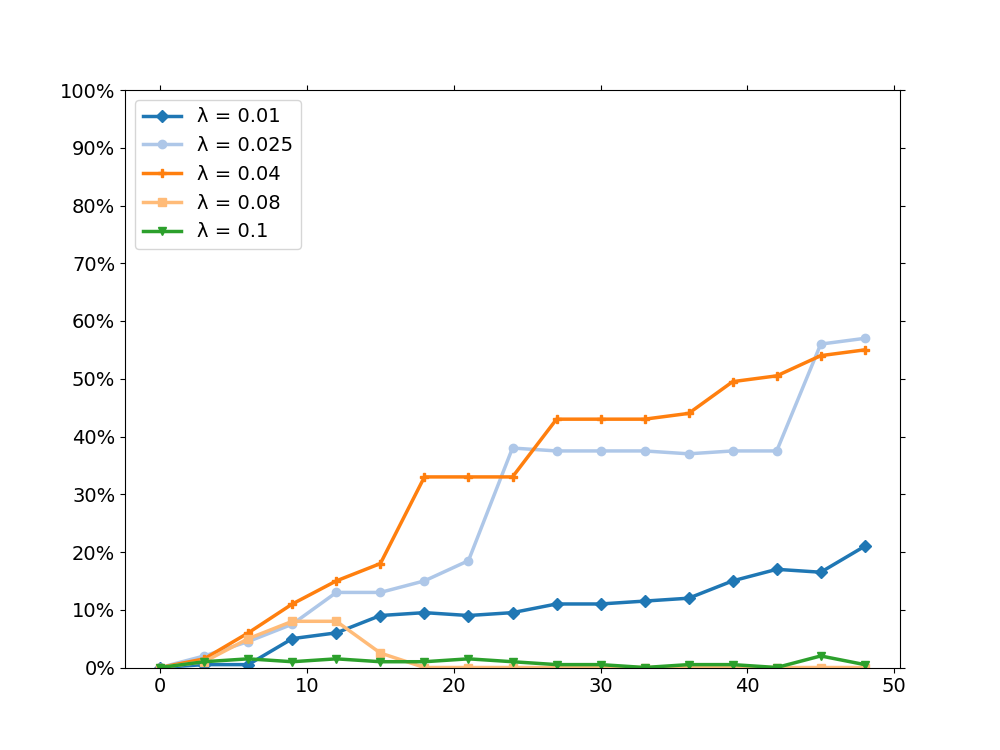}};
\node[below=of img1, node distance=0cm, yshift=1.5cm] {\scriptsize Normalised Optimiser Iterations};
\node[left=of img1, node distance=0cm, rotate=90, anchor=center,yshift=-1.5cm] {\scriptsize Prob. of Optimal Solution};
\node[right=of img1] (img2)  {\includegraphics[scale=0.32]{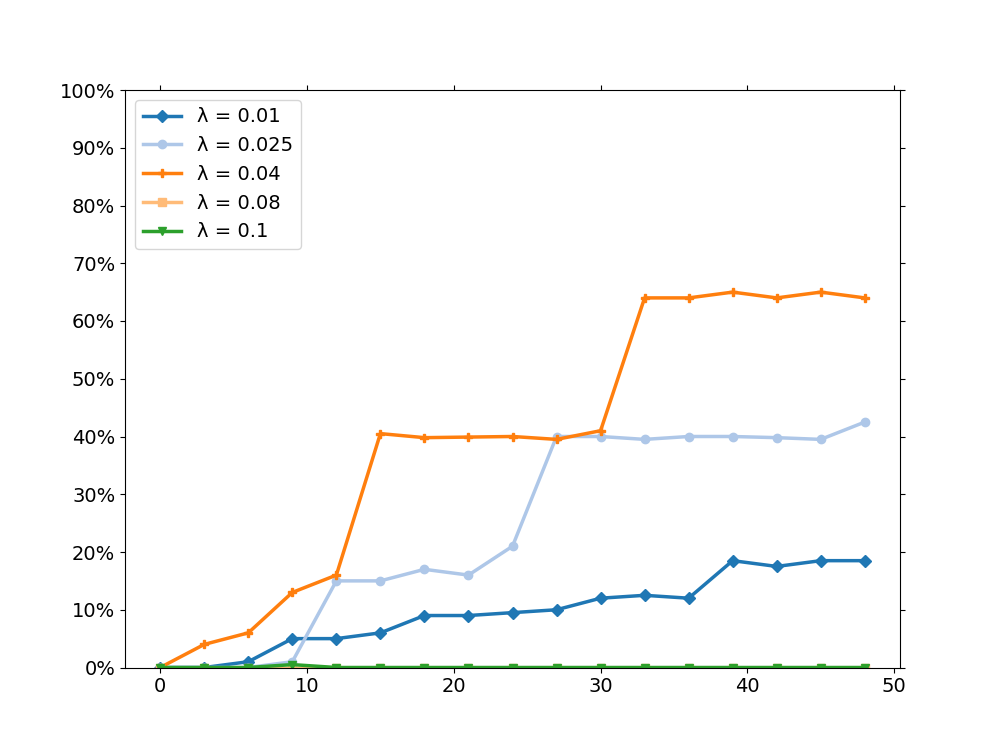}};
\node[below=of img2, node distance=0cm, yshift=1.5cm] {\scriptsize Normalised Optimiser Iterations};
\node[left=of img2, node distance=0cm, rotate=90, anchor=center,yshift=-1.5cm] {\scriptsize Prob. of Optimal Solution};
\end{tikzpicture}
\caption{Performance of \emph{Ascending-CVaR} algorithm with linear ascending for different choices of the ascending factor $\lambda$. The yellow (square marker) and green (down-pointing marker) lines which refer to $\alpha = 0.08$ and $\alpha = 0.1$ respectively are not able to reach a good approximation to the optimal solution. On the other hand, the orange (line marker) and light blue (circle marker) lines which correspond to $\alpha = 0.04$ and $\alpha=0.025$ are both able to achieve an overlap larger than $50\%$. Finally, the dark blue line (diamond marker) is still able to reach a good approximation to the ground state but it lacks in terms of speed of convergence and magnitude of overlap achieved. The graph on the left refers to a Number Partitioning problem while the graph on the right to  Portfolio Optimisation Problem.}
\label{fig:ascending_factor}
\end{figure*}
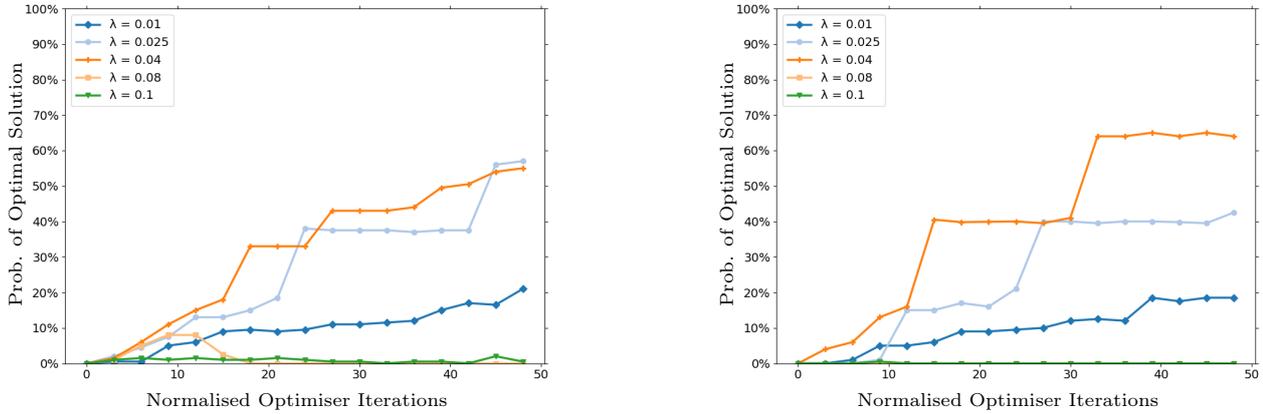

In this section we illustrate how the performance of our algorithm depends on the choice of the \emph{ascending factor} $\lambda$. We numerically tested a large number of instances of sizes from 16 to 20 qubits and observed that the algorithm performed optimally for ascending factors drawn from the set $[0.025, 0.045]$. For this reason, as an example, we choose to draw the behavior of our algorithm for two random instances (one for Portfolio Optimisation and one for Number Partitioning) for different choices of the hyperparameter $\lambda$.

The performance of our method is sensitive to the choice of $\lambda$. A small $\lambda$ still converges to an optimal solution but requires a large number of iterations, compared to $\lambda$ chosen from the set $[0.025, 0.045]$ which is able to attain a $10\%$ within a small number of iterations. On the other hand, choosing $\lambda$ to be large (hoping for a faster convergence) fails to achieve even a minor overlap with the optimal solution. A careful tuning of $\lambda$ is therefore necessary for the optimal performance of the algorithm given the size and class of the problem at hand.

\end{document}